\documentclass[
 amsmath,amssymb,
 aps,
 twocolumn,
 prd,
 showpacs,
 superscriptaddress, 
 floatfix
]{revtex4-2}

\usepackage{color}

\usepackage{comment}
\usepackage{graphicx}
\usepackage{dcolumn}
\usepackage{bm}
\usepackage{enumitem} 
\usepackage{todonotes}
\setlist{itemsep=0pt}
\usepackage{siunitx}
\usepackage{booktabs}

\usepackage{upgreek}

\usepackage{xspace}
\usepackage{hyperref}
\usepackage{nicefrac}
\hypersetup{
  colorlinks=true,
  linkcolor=black,
  filecolor=magenta,      
  urlcolor=blue,
}

\usepackage[
  noabbrev,
  capitalize
]{cleveref}

\newcommand{\aprime}{\ensuremath{A^\prime}\xspace}

\newcommand{\ele}{\ensuremath{e^-}\xspace}
\newcommand{\pos}{\ensuremath{e^+}\xspace}
\newcommand{\epem}{\pos{}\ele{}\xspace}

\newcommand{\eps}{\ensuremath{\epsilon{}}\xspace}
\newcommand{\radfrac}{\ensuremath{f_\text{rad}(m_{\aprime})}}
\newcommand{\radacc}{\ensuremath{A_\text{rad}(m_{\aprime})}}

\newcommand{\accxeff}{\ensuremath{\text{acceptance}\times\text{efficiency}}\xspace}

\newcommand{\Psum}{\ensuremath{P_\mathrm{sum}}\xspace}
\newcommand{\psum}{\ensuremath{p_\mathrm{sum}}\xspace}
\newcommand{\minyzero}{\ensuremath{y_{0,\min}}\xspace}

\newcommand{\vd}{\ensuremath{V_{D}}\xspace}

\newcommand{\rhod}{\ensuremath{\rho_{D}}\xspace}
\newcommand{\phid}{\ensuremath{\phi_{D}}\xspace}

\newcommand{\myfactors}{\frac{3\pi m_\aprime\epsilon^2}{2 N_{\text{eff}=1} \alpha}}

\usepackage[printonlyused]{acronym}
\acrodef{hps}[HPS]{Heavy Photon Search experiment}
\acrodef{jlab}[JLab]{Thomas Jefferson National Lab}
\acrodef{hs}[HS]{hidden sector}
\acrodef{bsm}[BSM]{beyond-standard-model}
\acrodef{simp}[SIMP]{strongly interacting massive particle}
\acrodef{dpi}[$\pi_D$]{dark pion}
\acrodefplural{dpi}[$\pi_D$]{dark pions}
\acrodef{dvec}[$V_D$]{dark vector meson}
\acrodefplural{dvec}[$V_D$]{dark vector mesons}
\acrodef{svt}[SVT]{silicon vertex tracker}
\acrodef{ecal}[ECal]{electromagnetic calorimeter}
\acrodef{sm}[SM]{Standard Model}
\acrodef{sr}[SR]{signal region}
\acrodef{cr}[CR]{control region}
\acrodef{mc}[MC]{Monte Carlo}
\acrodef{dm}[DM]{Dark Matter}
\acrodef{wzw}[WZW]{Wess-Zumino-Witten}
\acrodef{rad}[RAD]{Radiative Tridents}
\acrodef{bh}[BH]{Bethe-Heitler}
\acrodef{wab}[WAB]{wide-angle Bremsstrahlung}
\acrodef{slic}[SLIC]{Simulator for the Linear Collider}
\acrodef{ics}[ICS]{Integrated Cross Section}
\acrodef{kf}[KF]{Kalman Filter}
\acrodef{gbl}[GBL]{General Broken Lines}
\acrodef{cebaf}[CEBAF]{Continuous Electron Beam Accelerator Facility}
\acrodef{fee}[FEE]{full energy electron}
\acrodef{oim}[OIM]{Optimum Interval Method}
\acrodef{vps}[VPS]{vertex projection significance}

\usepackage{tikz}
\usetikzlibrary{%
  math,
  decorations.markings,
  decorations.pathreplacing,
  decorations.pathmorphing, 
  shapes.geometric,
  arrows.meta, 
}
\usetikzlibrary{fit} 
\usetikzlibrary{positioning, calc}
\usepackage{tkz-euclide}

\definecolor{HPSTarget}{rgb}{0.55,0.52,0.54}
\definecolor{HPSTracker}{rgb}{0.0,0.5,0.5}
\definecolor{HPSEcal}{rgb}{0.8,0.5,0.2}

\tikzmath{%
  \openingslope = 0.15; 
  \sensorsep = 0.12; 
  \sensorlen = 1.5; 
  \targetx = -3.0;
  \layeronex = -0.2;
  \layertwox = 2.0;
  \targethalfthick = 0.1;
  \ecalbarlen=1.5;
  \ecalbarwidth=0.3;
  \layeroney = (\layeronex - \targetx) * \openingslope;
  \layertwoy = (\layertwox - \targetx) * \openingslope;
  \sensorhalfsep = \sensorsep / 2;
  \reflineendx = \layertwox+0.5;
  \reflineendy = (\reflineendx - \targetx) * \openingslope;
  \eleslope=(2-0.1)/(2.5-\targetx-2);
  \eleyzero=\eleslope*(-2.0)+0.1;
  \elelayeronestereoy=\eleslope*(\layeronex - \sensorhalfsep+1.0)+0.1;
  \elelayeroneaxialy=\eleslope*(\layeronex + \sensorhalfsep+1.0)+0.1;
  \elelayertwostereoy=\eleslope*(\layertwox - \sensorhalfsep+1.0)+0.1;
  \elelayertwoaxialy=\eleslope*(\layertwox + \sensorhalfsep+1.0)+0.1;
  \posslope=(-1.8-0.1)/(2.5-\targetx-2);
  \posyzero=\posslope*(-2.0)+0.1;
  \poslayeronestereoy=\posslope*(\layeronex - \sensorhalfsep+1.0)+0.1;
  \poslayeroneaxialy=\posslope*(\layeronex + \sensorhalfsep+1.0)+0.1;
  \poslayertwostereoy=\posslope*(\layertwox - \sensorhalfsep+1.0)+0.1;
  \poslayertwoaxialy=\posslope*(\layertwox + \sensorhalfsep+1.0)+0.1;
  \ndeleslope = (2-0)/(2.5-\targetx-\targethalfthick);
  \ndelelayeronestereoy=\ndeleslope*(\layeronex - \sensorhalfsep - \targetx - \targethalfthick);
  \ndelelayeroneaxialy=\ndeleslope*(\layeronex + \sensorhalfsep - \targetx - \targethalfthick);
  \ndelelayertwostereoy=\ndeleslope*(\layertwox - \sensorhalfsep - \targetx - \targethalfthick);
  \ndelelayertwoaxialy=\ndeleslope*(\layertwox + \sensorhalfsep - \targetx - \targethalfthick);
  \ndposslope = (-1.8-0)/(2.5-\targetx-\targethalfthick);
  \ndposlayeronestereoy=\ndposslope*(\layeronex - \sensorhalfsep - \targetx - \targethalfthick);
  \ndposlayeroneaxialy=\ndposslope*(\layeronex + \sensorhalfsep - \targetx - \targethalfthick);
  \ndposlayertwostereoy=\ndposslope*(\layertwox - \sensorhalfsep - \targetx - \targethalfthick);
  \ndposlayertwoaxialy=\ndposslope*(\layertwox + \sensorhalfsep - \targetx - \targethalfthick);
  \fdeleslope = (2-0)/(2.5-\targetx-\targethalfthick);
  \fdelelayeronestereoy=\fdeleslope*(\layeronex - \sensorhalfsep - \targetx - \targethalfthick);
  \fdelelayeroneaxialy=\fdeleslope*(\layeronex + \sensorhalfsep - \targetx - \targethalfthick);
  \fdelelayertwostereoy=\fdeleslope*(\layertwox - \sensorhalfsep - \targetx - \targethalfthick);
  \fdelelayertwoaxialy=\fdeleslope*(\layertwox + \sensorhalfsep - \targetx - \targethalfthick);
  \fdposx = \targetx+1.0;
  \fdposy = \fdeleslope*(\fdposx-\targetx-\targethalfthick);
  \fdposslope = (-1.8-\fdposy)/(2.5-\fdposx);
  \fdposyzero = \fdposslope*(\targetx-\fdposx)+\fdposy;
  \fdbadhitstereoy = \fdposslope*(\layeronex-\sensorhalfsep-\fdposx)+\fdposy;
  \fdbadhitaxialy = \fdposslope*(\layeronex+\sensorhalfsep-\fdposx)+\fdposy;
  \fdposlayertwostereoy=\fdposslope*(\layertwox - \sensorhalfsep - \fdposx)+\fdposy;
  \fdposlayertwoaxialy=\fdposslope*(\layertwox + \sensorhalfsep - \fdposx)+\fdposy;
  \fdpostruslope = (\fdposslope*(\layertwox-\fdposx)+\fdposy-0)/(\layertwox-\targetx-\targethalfthick);
  \fdtruhitstereoy = \fdpostruslope*(\layeronex-\sensorhalfsep-\targetx-\targethalfthick);
  \fdtruhitaxialy = \fdpostruslope*(\layeronex+\sensorhalfsep-\targetx-\targethalfthick);
}

\tikzset{
  vertex/.style={diamond,fill,inner sep=1.5pt},
  hit/.style={circle,fill,inner sep=1.5pt},
  miss/.style={circle,draw,fill=white,inner sep=1.0pt}
}

\newcommand{\drawhpsfirsttwolayers}{%
  \draw[HPSTracker, very thick]
    (\layeronex-\sensorhalfsep,-\layeroney-\sensorlen)
    -- (\layeronex-\sensorhalfsep,-\layeroney)
    (\layeronex+\sensorhalfsep,-\layeroney-\sensorlen)
    -- (\layeronex+\sensorhalfsep,-\layeroney)
    (\layeronex-\sensorhalfsep,\layeroney+\sensorlen)
    -- (\layeronex-\sensorhalfsep,\layeroney)
    (\layeronex+\sensorhalfsep,\layeroney+\sensorlen)
    -- (\layeronex+\sensorhalfsep,\layeroney);
  \draw[HPSTracker] (\layeronex,\layeroney+\sensorlen) node[above] {L1};  
  \draw[HPSTracker, very thick]
    (\layertwox-\sensorhalfsep,-\layertwoy-\sensorlen)
    -- (\layertwox-\sensorhalfsep,-\layertwoy)
    (\layertwox+\sensorhalfsep,-\layertwoy-\sensorlen)
    -- (\layertwox+\sensorhalfsep,-\layertwoy)
    (\layertwox-\sensorhalfsep,\layertwoy+\sensorlen)
    -- (\layertwox-\sensorhalfsep,\layertwoy)
    (\layertwox+\sensorhalfsep,\layertwoy+\sensorlen)
    -- (\layertwox+\sensorhalfsep,\layertwoy);
  \draw[HPSTracker] (\layertwox,\layertwoy+\sensorlen) node[above] {L2};
  \draw[gray,dashed] (\targetx,0) -- (\reflineendx,\reflineendy);
  \draw[gray,dashed] (\targetx,0) -- (\reflineendx,0);
  \draw[gray,dashed] (\targetx,0) -- (\reflineendx,-\reflineendy);
  \fill[HPSTarget] (\targetx-\targethalfthick,+1) rectangle (\targetx+\targethalfthick,-1);
  \draw[HPSTarget] (\targetx,-1) node[below] {Target};
}

\begin{document}

\preprint{SLAC/XXX-XXX}

\title{%
    First Displaced Vertex
Search for Electroproduced Dark-Sector Strongly Interacting Massive Particles by the
HPS Experiment
}

\date{\today}

\collaboration{HPS Collaboration}\noaffiliation

\begin{abstract}
  The Heavy Photon Search experiment (HPS) is a fixed-target, electron beam experiment designed to search for  $e^+e^-$ mass resonances and displaced decays using a forward acceptance spectrometer. 
This article details the search for naturally long-lived ``dark" vector mesons ($V_D$) arising from a dark sector of beyond-Standard-Model \acp{simp}, characterized by a QCD-like $SU(3)_D$ symmetry and coupled to the Standard Model photon via a new $U(1)_D$ gauge interaction mediated by the ``heavy photon", or $A^\prime$.
The results are based on an integrated luminosity of \SI{10608}{nb^{-1}} collected during the 2016 HPS Engineering Run.
The displaced vertex search for $V_D \rightarrow e^+e^-$ in the $e^+e^-$ invariant mass range \SIrange[]{39}{179}{MeV} showed no statistically significant evidence for signal above the QED background.

\end{abstract}

%
%
%
\newcommand*{\SACLAY}{IRFU, CEA, Universit\'e Paris-Saclay, F-91190 Gif-sur-Yvette, France}
\newcommand*{\SACLAYindex}{1}
\newcommand*{\INFNGE}{INFN, Sezione di Genova, 16146 Genova, Italy}
\newcommand*{\INFNGEindex}{2}
\newcommand*{\INFNTUR}{INFN, Sezione di Torino, 10125 Torino, Italy}
\newcommand*{\INFNTURindex}{21}
\newcommand*{\INFNMB}{University di Milano Bicocca, 20126 Milano, Italy}
\newcommand*{\INFNMBindex}{56}
\newcommand*{\ORSAY}{Université Paris-Saclay, CNRS, IJCLab, 91405, Orsay, France}
\newcommand*{\ORSAYindex}{22}
\newcommand*{\UNH}{University of New Hampshire, Durham, New Hampshire 03824, USA}
\newcommand*{\UNHindex}{27}
\newcommand*{\NSU}{Norfolk State University, Norfolk, Virginia 23504, USA}
\newcommand*{\NSUindex}{28}
\newcommand*{\ODU}{Old Dominion University, Norfolk, Virginia 23529, USA}
\newcommand*{\ODUindex}{30}
\newcommand*{\ROMAII}{Universit\`a di Roma Tor Vergata, 00133 Rome Italy}
\newcommand*{\ROMAIIindex}{32}
\newcommand*{\JLAB}{Thomas Jefferson National Accelerator Facility, Newport News, Virginia 23606, USA}
\newcommand*{\JLABindex}{36}
\newcommand*{\GLASGOW}{University of Glasgow, Glasgow G12 8QQ, United Kingdom}
\newcommand*{\GLASGOWindex}{39}
\newcommand*{\WM}{College of William \& Mary, Williamsburg, Virginia 23187, USA}
\newcommand*{\WMindex}{42}
\newcommand*{\YEREVAN}{Yerevan Physics Institute, 375036 Yerevan, Armenia}
\newcommand*{\YEREVANindex}{43}
\newcommand*{\PERIMETER}{Perimeter Institute, Ontario, Canada N2L 2Y5}
\newcommand*{\PERIMETERindex}{44}
\newcommand*{\SASSARI}{Universit\`a di Sassari, 07100 Sassari, Italy}
\newcommand*{\SASSARIindex}{45}
\newcommand*{\FNAL}{Fermi National Accelerator Laboratory, Batavia, IL 60510, USA}
\newcommand*{\FNALindex}{46} 
\newcommand*{\CATANIA}{INFN, Sezione di Catania, 95123 Catania, Italy}
\newcommand*{\CATANIAindex}{47}
\newcommand*{\STONYBROOK}{C.~N.~Yang Institute for Theoretical Physics, 
                          Stony Brook University, Stony Brook, NY 11794, USA}
\newcommand*{\STONYBROOKindex}{48}
\newcommand*{\SLAC}{SLAC National Accelerator Laboratory, Stanford University, Stanford, CA 94309, USA}
\newcommand*{\SLACindex}{49}
\newcommand*{\UCSC}{Santa Cruz Institute for Particle Physics, University of California, Santa Cruz, CA 95064, USA}
\newcommand*{\UCSCindex}{50}
\newcommand*{\PADOVA}{Universit\`a di Padova, 35122 Padova, Italy} 
\newcommand*{\PADOVAindex}{51}
\newcommand*{\INFNPA}{INFN, Sezione di Padova, 16146 Padova, Italy}
\newcommand*{\INFNPAindex}{52}
\newcommand*{\IDAHO}{Idaho State University, Pocatello, ID, 83209, USA}
\newcommand*{\IDAHOindex}{53}
\newcommand*{\INFNROMA}{INFN, Sezione di Roma Tor Vergata, 00133 Rome, Italy}
\newcommand*{\INFNROMAindex}{54}
\newcommand*{\INFNSUD}{INFN, Laboratori Nazionali del Sud, 95123 Catania, Italy}
\newcommand*{\INFNSUDindex}{55}
\newcommand*{\UMN}{University of Minnesota, Minneapolis, MN 55455, USA}
\newcommand*{\UMNindex}{56}
\newcommand*{\STANFORD}{Stanford University, Stanford, CA 94305, USA}
\newcommand*{\STANFORDindex}{57}
\newcommand*{\UCATANIA}{Università degli Studi di Catania, Dipartimento di Fisica e Astronomia “Ettore Majorana”, 95123 Catania, Italy}
\newcommand*{\UCATANIAindex}{58}
\newcommand*{\INFNCAG}{INFN, Sezione di Cagliari, 09042 Monserrato, Italy}
\newcommand*{\INFNCAGindex}{59}
\newcommand*{\TORONTO}{University of Toronto Department of Physics, Toronto, Ontario M5S1A7 Canada}
\newcommand*{\TORONTOindex}{60}
\newcommand*{\FIU}{Florida International University, Miami, FL 33199, USA}
\newcommand*{\FIUindex}{61}
\newcommand*{\CATHOLICU}{Catholic University of America, Washington, DC 20064, USA}
\newcommand*{\CATHOLICUindex}{61}

\author{P.~H.~Adrian}\affiliation\SLAC
\author{N.~A.~Baltzell}\affiliation\JLAB
\author{M.~Battaglieri}\affiliation\INFNGE
\author{M.~Bond\'i}\affiliation\CATANIA
\author{S.~Boyarinov}\affiliation\JLAB
\author{C.~Bravo}\affiliation\SLAC
\author{S.~Bueltmann}\affiliation\ODU
\author{P.~Butti}\affiliation\SLAC
\author{V.~D.~Burkert}\affiliation\JLAB
\author{D.~Calvo}\affiliation\INFNTUR
\author{T.~Cao}\affiliation\UNH\affiliation\JLAB
\author{M.~Carpinelli}\affiliation\INFNMB\affiliation\INFNSUD
\author{A.~Celentano}\affiliation\INFNGE
\author{G.~Charles}\affiliation\ORSAY
\author{L.~Colaneri}\affiliation\ROMAII\affiliation\INFNROMA
\author{W.~Cooper}\affiliation\FNAL
\author{B.~Crowe}\affiliation\UNH
\author{C.~Cuevas}\affiliation\JLAB
\author{A.~D'Angelo}\affiliation\ROMAII\affiliation\INFNROMA
\author{N.~Dashyan}\affiliation\YEREVAN
\author{M.~De~Napoli}\affiliation\UCATANIA\affiliation\CATANIA
\author{R.~De~Vita}\affiliation\INFNGE\affiliation\JLAB
\author{A.~Deur}\affiliation\JLAB
\author{M.~Diamond}\affiliation\SLAC\affiliation\TORONTO
\author{R.~Dupre}\affiliation\ORSAY
\author{H.~Egiyan}\affiliation\JLAB
\author{T.~Eichlersmith}\affiliation\UMN
\author{L.~Elouadrhiri}\affiliation\JLAB
\author{R.~Essig}\affiliation\STONYBROOK
\author{V.~Fadeyev}\affiliation\UCSC
\author{C.~Field}\affiliation\SLAC
\author{A.~Filippi}\affiliation\INFNTUR
\author{A.~Freyberger}\affiliation\JLAB
\author{S.~Gaiser}\affiliation{\STANFORD}
\author{M.~Gar\c{c}on}\affiliation\SACLAY
\author{N.~Gevorgyan}\affiliation\YEREVAN
\author{F.~X.~Girod}\affiliation\JLAB
\author{N.~Graf}\affiliation\SLAC
\author{M.~Graham}\thanks{Corresponding Author: mgraham@slac.stanford.edu}\affiliation\SLAC
\author{K.~A.~Griffioen}\affiliation\WM
\author{A.~Grillo}\affiliation\UCSC
\author{M.~Guidal}\affiliation\ORSAY
\author{R.~Herbst}\affiliation\SLAC
\author{M.~Holtrop}\affiliation\UNH
\author{J.~Jaros}\affiliation\SLAC
\author{R.~P.~Johnson}\affiliation\UCSC
\author{G.~Kalicy}\affiliation\CATHOLICU
\author{M.~Khandaker}\affiliation\IDAHO
\author{V.~Kubarovsky}\affiliation\JLAB
\author{E.~Leonora}\affiliation\CATANIA
\author{K.~Livingston}\affiliation\GLASGOW
\author{L.~Marsicano}\affiliation\INFNGE
\author{T.~Maruyama}\thanks{Deceased}\affiliation\SLAC
\author{S.~McCarty}\affiliation\UNH
\author{J.~McCormick}\affiliation\SLAC
\author{B.~McKinnon}\affiliation\GLASGOW
\author{K.~Moffeit}\affiliation\GLASGOW
\author{O.~Moreno}\affiliation\SLAC\affiliation\UCSC
\author{C.~Munoz~Camacho}\affiliation\ORSAY
\author{T.~Nelson}\affiliation\SLAC
\author{S.~Niccolai}\affiliation\ORSAY
\author{A.~Odian}\affiliation\SLAC
\author{R.~O'Dwyer}\affiliation\STANFORD
\author{M.~Oriunno}\affiliation\SLAC
\author{M.~Osipenko}\affiliation\INFNGE
\author{R.~Paremuzyan}\affiliation\JLAB
\author{S.~Paul}\affiliation\FIU
\author{E.~Peets}\affiliation{\STANFORD}
\author{N.~Randazzo}\affiliation\CATANIA
\author{B.~Raydo}\affiliation\JLAB
\author{B.~Reese}\affiliation\SLAC
\author{A.~Rizzo}\affiliation\ROMAII\affiliation\INFNROMA
\author{P.~Schuster}\affiliation\SLAC\affiliation\PERIMETER
\author{Y.~G.~Sharabian}\affiliation\JLAB
\author{G.~Simi}\affiliation\PADOVA\affiliation\INFNPA
\author{A.~Simonyan}\affiliation\ORSAY
\author{V.~Sipala}\affiliation\SASSARI\affiliation\INFNCAG
\author{A.~Spellman}\affiliation\UCSC
\author{D.~Sokhan}\affiliation\GLASGOW
\author{M.~Solt}\affiliation\SLAC
\author{S.~Stepanyan}\affiliation\JLAB
\author{H.~Szumila-Vance}\affiliation\FIU
\author{L.~Tompkins}\affiliation{\STANFORD}
\author{N.~Toro}\affiliation\SLAC\affiliation\PERIMETER
\author{S.~Uemura}\affiliation\SLAC
\author{M.~Ungaro}\affiliation\JLAB
\author{H.~Voskanyan}\affiliation\YEREVAN
\author{L.~B.~Weinstein}\affiliation\ODU
\author{B.~Wojtsekhowski}\affiliation\JLAB

\rule{0pt}{0pt}

\maketitle
\section{Introduction}\label{sec:introduction}

In recent years, a number of extensions to the \ac{sm} have been proposed which include new gauge symmetries that allow for so-called dark sectors with indirect coupling to the \ac{sm} to account for the dark matter~\cite{Hewett:2012ns,Essig:2013lka,Alexander:2016aln,Battaglieri:2017aum}. In the simplest of these, a new $U(1)_D$ gauge field is introduced in the dark sector, giving rise to a potentially massive spin-1 vector gauge boson referred to as the ``dark photon", or \aprime~\cite{Holdom:1985ag,Galison:1983pa,Fayet:1990wx}.
The dark photon kinetically mixes with the \ac{sm} photon through a massive charged fermion loop, a process that is often simplified to an effective coupling with strength $\epsilon$.
This coupling enables the electro-production of dark photons through a bremsstrahlung-like process on a nuclear target~\cite{AprimeFixedTargetTheory}.

The final state signatures from the dark photon decay depend on the structure of the dark sector. Our previous analyses \cite{Adrian:2022nkt, Adrian:2018scb} were optimized to search for an \aprime in the simplest case where the \aprime, being light compared to other dark states, can only decay back into \ac{sm} leptons.  There are a number of other models in the literature, some of which will give different signatures in the \ac{hps} detector. In this work, we present a search for particles predicted by the Strongly Interacting Massive Particles (\ac{simp}) model~\cite{PhysRevLett.113.171301,PhysRevLett.115.021301}.  
In \Cref{sec:model}, this paper discusses the \ac{simp} model, highlighting both theoretical and experimental constraints. This is followed by brief descriptions of the \ac{hps} experiment in \Cref{sec:experiment}, and the data collection and reconstruction in \Cref{sec:dataset}. \Cref{sec:selection} and \Cref{sec:analysis} detail the event selection and data analysis, respectively.  \Cref{sec:conclusion} summarizes the findings and suggests possible improvements for future analyses. 

\section{SIMP Model and Parameter Constraints}
\label{sec:model}
\begin{figure}[ht]
    \centering
    \resizebox{0.45\textwidth}{!}{\tikzset{snake it/.style={decorate, decoration=snake}}
\tikzset{
  on each segment/.style={
    decorate,
    decoration={
      show path construction,
      moveto code={},
      lineto code={
        \path [#1]
        (\tikzinputsegmentfirst) -- (\tikzinputsegmentlast);
      },
      curveto code={
        \path [#1] (\tikzinputsegmentfirst)
        .. controls
        (\tikzinputsegmentsupporta) and (\tikzinputsegmentsupportb)
        ..
        (\tikzinputsegmentlast);
      },
      closepath code={
        \path [#1]
        (\tikzinputsegmentfirst) -- (\tikzinputsegmentlast);
      },
    },
  },
  mid arrow/.style={thick,postaction={decorate,decoration={
        markings,
        mark=at position .5 with {\arrow[#1]{to}}
      }}},
}
\usetikzlibrary{fit} 

\tikzmath{%
  \startgammax = 0.85;
  \startgammay = 0.5;
  \gammalength = 1.0;
  %
}

\begin{tikzpicture}
    \draw[thick, postaction={on each segment={mid arrow}}] (-1.5,-1.) -- (0.75,-1.);
    \node[anchor=east] at (-1.5,-1.) (zin) {$Z$};    
    \draw[thick, postaction={on each segment={mid arrow}}] (-1.5,-0.2) -- (0.75,-0.2);
    \node[anchor=east] at (-1.5,-0.2) (elein) {$e^-$};    
    \draw[thick, snake it] (0.2,-1.) -- (0.2,-0.2);
    \node[anchor=west] at (0.2,-0.6) (gamma) {$\gamma$};    
    \draw[thick, snake it] (-1.1,-0.2) -- (-0.5,0.6) node [midway, above left] (Ap) {\aprime};    
    \draw[thick, double] (-0.5,0.6) -- (0.3,1.0) node [midway, above] {$V_D$};
    \draw[thick, dashed] (-0.5,0.6) -- (0.3,0.2) node [right] {$\pi_D$};
    \node[circle,fill,inner sep=1.25pt] at (-0.5,0.6) (apdecay) {};
    \node[anchor=east] at (-1.,1.2) (prompt) {prompt};
    \draw[->, red] (prompt.east) -- (-0.55,0.65);
    \node[anchor=west] at (1,-0.2) (missE) {missing energy};
    \draw[->, red] (missE.west) -- (0.7,0.05);
    \draw[dotted, thick] (0.3,1.0) -- (0.7,1.2);
    \draw[thick, double] (0.7,1.2) -- (1.5,1.6) node [midway, above] {$V_D$};
    \node[circle,fill,inner sep=1.25pt] at (1.5,1.6) (vddecay) {};
    \node[anchor=west] at (1.,0.6) (displaced) {displaced};
    \draw[->, red] (displaced.west) -- (0.55,1.1);
    \draw[thick, postaction={on each segment={mid arrow}}] (1.5,1.6) -- (2.1, 1.2);
    \node[anchor=west] at (2.1, 1.2) (ele2) {$e^{-}$};
    \draw[thick, postaction={on each segment={mid arrow}}] (2.1, 2.) -- (1.5,1.6);
    \node[anchor=west] at (2.1, 2.) (pos2) {$e^{+}$};
    \draw[thick] (-3.8,-0.2) -- (-3.2,-0.2);
    \node[anchor=west] at (-3.2,-0.2) (mpid) {$\pi_D$};
    \draw[thick] (-3.8,0.2) -- (-3.2,0.2);
    \node[anchor=west] at (-3.2,0.2) (mvd) {$V_D$};
    \draw[thick] (-3.8,0.9) -- (-3.2,0.9);
    \node[anchor=west] at (-3.2,0.9) (map) {\aprime};
    \draw[->] (-4.2,-0.2) -- (-4.2,1.2);
    \node[anchor=south] at (-4.2,1.2) (mass) {mass};
    \node[draw, thick, rectangle, gray, fit=(mass) (mpid) (map)] (fit) {};
\end{tikzpicture}}
    \caption{Production of $e^{+}e^{-}$ from the decay of a dark vector meson $V_D$ via a virtual dark photon \aprime.}
    \label{fig:intro:darkdecay}
\end{figure}
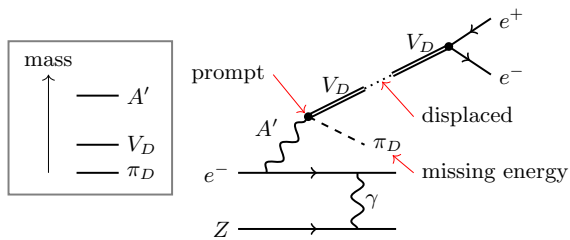

In contrast to the minimal dark photon model, where thermal freeze-out is achieved through $2 \to 2$ annihilation into \ac{sm} particles, extended dark sector models permit alternative freeze-out mechanisms. 
Introducing QCD-like $SU(3)_D$ gauge symmetries in the hidden sector gives rise to \aclp{simp}, namely \acp{dpi} and \acp{dvec}, where the lightest states, the dark pions, serve as dark matter candidates.

While these models still require kinetic equilibration with the \ac{sm} to produce the relic abundance, dark pion self-interactions allow for an additional $3\pi_D \to 2\pi_D$ annihilation process that depletes the dark matter relic density even after decoupling from the \ac{sm}~\cite{PhysRevLett.113.171301}. The inclusion of \acp{dvec} further enables a semi-annihilation channel, $\pi_D \pi_D \to \pi_D V_D$, followed by the decay $V_D \to \text{SM}$ particles through a virtual \aprime. This decay can produce a displaced $e^+e^-$ pair (\Cref{fig:intro:darkdecay}), a signature well matched to the capabilities of the \ac{hps}  detector~\cite{simp-pheno-2018}. 

The \ac{simp} model considered in this paper involves six parameters: the dark photon, dark pion, and dark vector masses, $m_{\aprime}$, $m_{\pi_{D}}$, and $m_{V_{D}}$, respectively; the $\aprime$ kinetic mixing strength $\epsilon$ with the \ac{sm} photon; the hidden sector $U(1)_{D}$ gauge coupling constant $\alpha_{D}$; and finally, the ratio of the dark pion mass to the dark pion decay constant $m_{\pi_{D}}/f_{\pi_{D}}$.
These parameters are constrained by both theoretical consistency and experimental requirements. Perturbativity demands $\alpha_D < 1$ and  in this work $\alpha_D$ is fixed at $10^{-2}$. This implies $m_{\pi_{D}} / f_{\pi_{D}} \lesssim 4\pi$, since $m_{\pi_{D}} / f_{\pi_{D}} \sim g_D \sim 4\pi \alpha_D$.
The kinetic mixing parameter must fall within $10^{-6}<\epsilon<10^{-2}$~\cite{AprimeFixedTargetTheory}. Values of $\epsilon \gtrsim 10^{-2}$ suppress semi-annihilation, while $\epsilon \lesssim 10^{-6}$ fail to maintain kinetic equilibrium between the dark and visible sectors in the early universe~\cite{simp-pheno-2018}.

We search the parameter space for decays that are visible and reconstructible in the \ac{hps} detector; this yields constraints on the search:
\begin{itemize}
    \item $m_{\aprime} > 2m_{\pi_{D}}$ to suppress $\pi\pi \to \aprime\pi$
    \item $m_{\aprime} > m_{\pi_{D}} + m_{V_{D}}$ to allow $\aprime \to \pi_D V_D$
    \item $m_{\aprime} < 2m_\mu$ and $m_{\aprime} < 2m_{V_{D}}$ to favor decays with good acceptance in the detector
    \item $m_{V_{D}} < 2m_{\pi_{D}}$ to prevent $V_D \to \pi_D \pi_D$ and ensure visible decay
\end{itemize}

  To manage the complexity of the parameter space, a benchmark model with fixed mass ratios used in reference~ \cite{simp-pheno-2018} is adopted. The search is then performed as a function of $m_{\aprime}$ and $\epsilon$, for the representative value of $m_{\pi_{D}} / f_{\pi_{D}} = 4\pi$.

\section{The HPS Experiment}\label{sec:experiment}

\begin{figure}
    \centering
    \includegraphics[width=0.45\textwidth]{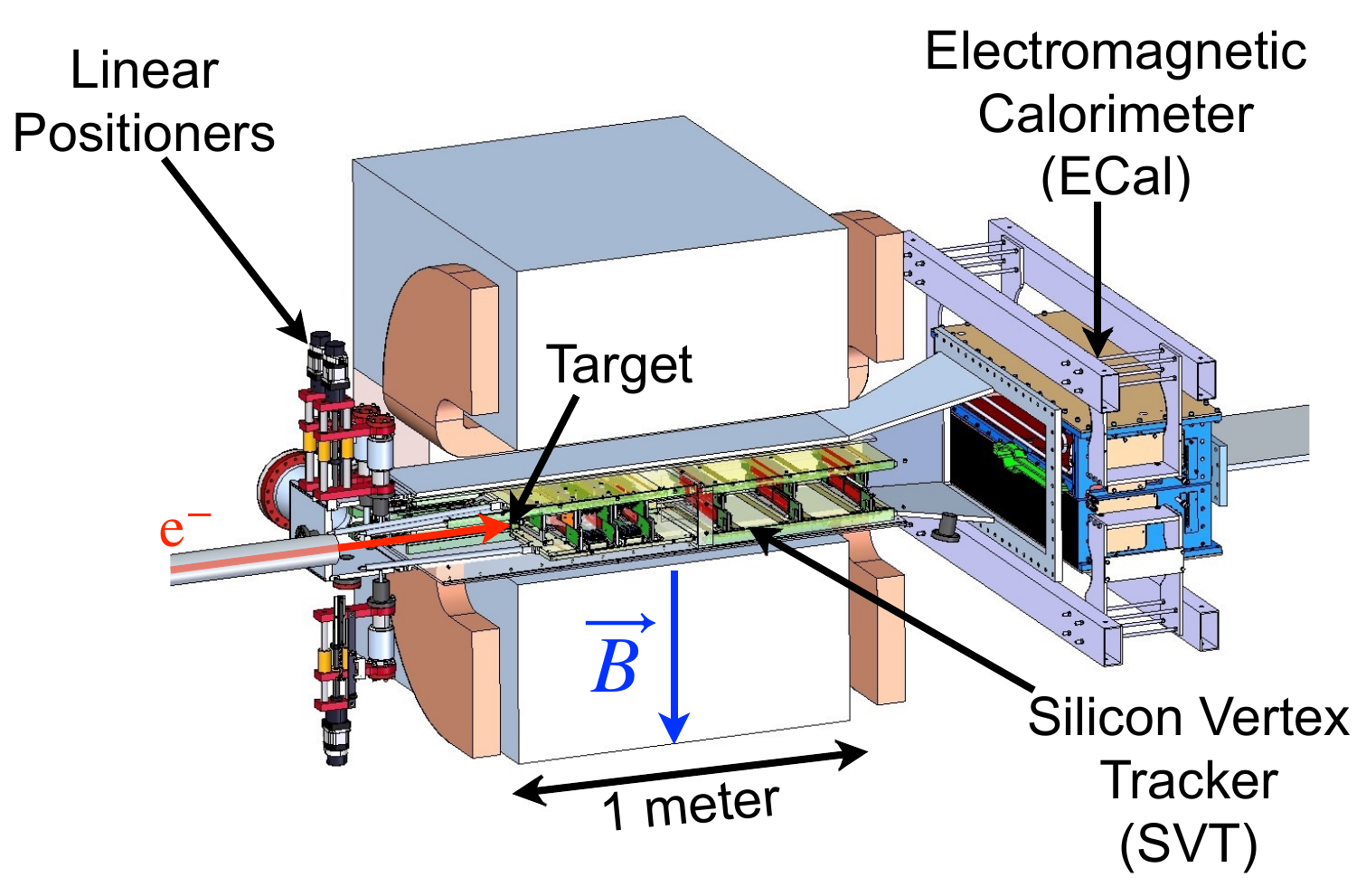}
    \caption{A cutaway view of the HPS detector showing the \ac{svt} in a vacuum chamber inside the bore of the spectrometer magnet and the downstream \ac{ecal}. The positions of the target and the front portions of the \ac{svt} are controlled by a set of linear positioning motors upstream of the detector.}
    \label{fig:hps_generalView}
\end{figure}

 This section provides an overview of the \ac{cebaf} accelerator and the \ac{hps} detector.  The key components of the \ac{hps} apparatus are shown in \cref{fig:hps_generalView}. More detailed motivations and detector specifications are discussed in \cite{Adrian:2022nkt}.

\Ac{hps} uses the electron beam from the \ac{cebaf}~\cite{Leemann:2001dg} at Thomas Jefferson National Accelerator Facility in Newport News, Virginia. 
\ac{cebaf}’s ability to provide a high-repetition-rate, multi-\si{GeV} electron beam with low per-bunch charge is essential to \ac{hps}, allowing for high-luminosity operation with minimal pile-up and manageable detector occupancies~\cite{HPS:2016jta}.

Although the \ac{hps} detector was designed to search for prompt and displaced \aprime, it is also sensitive to a subset of \ac{simp} decays, that produce similar $e^+e^-$ final states but with different kinematics. \ac{hps} targets rare $e^+e^-$ decays while rejecting large QED backgrounds. This requires a precise measurement of the invariant mass and the position of the decay vertex. The overall geometry of the detector is optimized for  forward-going $e^+e^-$ pairs, a characteristic shared by many potential signals, including both \aprime and SIMP decays. In the nominal \aprime scenario, the signal (and hence the $e^+e^-$ pair) carries nearly all the beam energy, peaking at $x = E_{\aprime}/E_\text{beam} \to 1$~\cite{AprimeFixedTargetTheory}. In contrast, for the \ac{simp} model, the \aprime decays to dark-sector particles which could be stable and leave no trace in the detector.   The channel of interest for this study, $\aprime \to \pi_D V_D$,  the $V_D$ decays to $e^+e^-$ while the $\pi_D$ escapes the detector and leads to missing momentum.  This results in lower $x$ for the pair and a less boosted decay with wider opening angles~\cite{simp-pheno-2018}. Although \ac{hps} has limited acceptance for such events, it remains sensitive in regions where the SIMP decay products fall within the detector’s forward coverage.

To produce forward $e^+e^-$ pairs, \ac{hps} places a thin (\SI{4}{\um}) tungsten foil target and Silicon Vertex Tracker (\ac{svt}) inside a dipole magnet. The magnetic field, with a magnitude of \SI{0.5}{T} for the 2016 run, bends charged particles in the horizontal ``beam plane''.  This separates electron from positron tracks and lower momentum signal tracks from beam-related backgrounds, mostly full-energy electrons or very low-momentum charged particles from the target.

The \ac{svt} is split into upper and lower halves, positioned just above and below the beam plane, to maximize acceptance near the beam while avoiding the large rate of scattered beam electrons. The \ac{svt} halves are placed at a vertical angle of approximately $\pm \qty{15}{mrad}$ from the beam plane. 
Each \ac{svt} half includes six modules of axial/stereo sensor pairs, arranged from 10 to 90 cm downstream of the target leading to a maximum number of measurements on a track of 12.  Each sensor has a \qty{60}{\micro m} readout strip pitch.  Strips are read out using APV25~\cite{French:2001xb} ASICs which records 6 samples of the signal development, allowing reconstruction of hit time with $\approx \SI{2}{ns}$ resolution.

The Electromagnetic Calorimeter (\ac{ecal})~\cite{Balossino:2016nly} sits downstream of the SVT. It is composed of 442 $\mathrm{PbWO_{4}}$ crystals arranged in two identical arrays above and below the beam plane. The \ac{ecal} serves two roles in the \ac{hps} experiment. First, it is used in the fast $e^+e^-$ trigger system, selecting events that have two clusters in opposite quadrants of the \ac{ecal}, i.e. in the top right and bottom left of the \ac{ecal} or vice versa. A detailed description of this trigger setup, referred to as Pair1 trigger, is given in~\cite{Adrian:2022nkt}. Second, it is used in particle reconstruction where we match the \ac{svt} track to an \ac{ecal} cluster helping to reduce background events from mis-reconstructed and out-of-time tracks.

\section{Data and Reconstruction} \label{sec:dataset}
The results presented here use data collected during the 2016 Engineering Run. All data used for analysis were collected at a beam energy of \SI{2.3}{GeV} with a current of \SI{200}{nA} on a tungsten foil target \SI{4}{\um} ($\approx$0.125\%~$X_0$) thick. The total luminosity of this dataset is \SI{10608}{nb^{-1}}, comprising 7.2 billion triggered events from a total charge on target of \SI{67.2}{mC}. In addition to physics runs, a number of special runs were taken, such as magnetic field-off runs and runs with a trigger dedicated to collecting scattered single electrons over a wide range of scattering angles. Data from these runs were used to calibrate and align the \ac{ecal} and \ac{svt} detectors. 

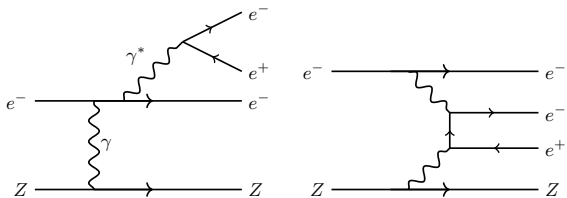
\begin{figure}[ht]
    \centering
    \resizebox{0.22\textwidth}{!}{\tikzset{snake it/.style={decorate, decoration=snake}}
\tikzset{
  on each segment/.style={
    decorate,
    decoration={
      show path construction,
      moveto code={},
      lineto code={
        \path [#1]
        (\tikzinputsegmentfirst) -- (\tikzinputsegmentlast);
      },
      curveto code={
        \path [#1] (\tikzinputsegmentfirst)
        .. controls
        (\tikzinputsegmentsupporta) and (\tikzinputsegmentsupportb)
        ..
        (\tikzinputsegmentlast);
      },
      closepath code={
        \path [#1]
        (\tikzinputsegmentfirst) -- (\tikzinputsegmentlast);
      },
    },
  },
  mid arrow/.style={thick,postaction={decorate,decoration={
        markings,
        mark=at position .5 with {\arrow[#1]{to}}
      }}},
}

\begin{tikzpicture}
\draw[thick] (-1.5,-1.5) -- (2,-1.5);
\draw[thick,->] (-0.5,-1.5) -- (0.5,-1.5);
\node[anchor=east] at (-1.5,-1.5) (zin) {$Z$};
\node[anchor=west] at (2,-1.5) (zout) {$Z$};
\draw[thick] (-1.5,0) -- (2,0);
\draw[thick,->] (-0.5,0) -- (0.5,0);
\node[anchor=east] at (-1.5,0) (elein) {$e^-$};
\node[anchor=west] at (2,0) (eleout) {$e^-$};
\draw[thick, snake it] (-0.5,-1.5) -- (-0.5,0);
\node[anchor=west] at (-0.5,-0.75) (gamma) {$\gamma$};
\draw[thick, snake it] (0,0) -- (1,1);
\node[anchor=south east] at (0.5,0.5) (gammabrems) {$\gamma^*$};
\draw[thick, postaction={on each segment={mid arrow}}] (2,0.5) -- (1,1);
\node[anchor=west] at (2,0.5) (lplus) {$e^+$};
\draw[thick, postaction={on each segment={mid arrow}}] (1,1) -- (2,1.5);
\node[anchor=west] at (2,1.5) (lminus) {$e^-$};
\end{tikzpicture}}%
    \resizebox{0.22\textwidth}{!}{\tikzset{snake it/.style={decorate, decoration=snake}}
\tikzset{
  on each segment/.style={
    decorate,
    decoration={
      show path construction,
      moveto code={},
      lineto code={
        \path [#1]
        (\tikzinputsegmentfirst) -- (\tikzinputsegmentlast);
      },
      curveto code={
        \path [#1] (\tikzinputsegmentfirst)
        .. controls
        (\tikzinputsegmentsupporta) and (\tikzinputsegmentsupportb)
        ..
        (\tikzinputsegmentlast);
      },
      closepath code={
        \path [#1]
        (\tikzinputsegmentfirst) -- (\tikzinputsegmentlast);
      },
    },
  },
  mid arrow/.style={thick,postaction={decorate,decoration={
        markings,
        mark=at position .5 with {\arrow[#1]{to}}
      }}},
}

\begin{tikzpicture}
\draw[thick] (-1.5,0) -- (2,0);
\draw[thick,->] (-0.5,0) -- (0.5,0);
\node[anchor=east] at (-1.5,0) (elein) {$e^-$};
\node[anchor=west] at (2,0) (eleout) {$e^-$};
\draw[thick] (-1.5,-2) -- (2,-2);
\draw[thick,->] (-0.5,-2) -- (0.5,-2);
\node[anchor=east] at (-1.5,-2) (zin) {$Z$};
\node[anchor=west] at (2,-2) (zout) {$Z$};
\draw[thick, snake it] (-0.2,-2) -- (0.5,-1.3);
\draw[thick, snake it] (-0.2,0) -- (0.5,-0.7);
\draw[thick, postaction={on each segment={mid arrow}}] (0.5,-1.3) -- (0.5,-0.7);
\draw[thick, postaction={on each segment={mid arrow}}] (2,-1.3) -- (0.5,-1.3);
\node[anchor=west] at (2,-1.3) (lplus) {$e^+$};
\draw[thick, postaction={on each segment={mid arrow}}] (0.5,-0.7) -- (2,-0.7);
\node[anchor=west] at (2,-0.7) (lminus) {$e^-$};
\end{tikzpicture}}
    \caption{%
      Radiative (left) and Bethe-Heitler tridents (right) have the same final state particles as the \epem production from a dark vector decay shown in \cref{fig:intro:darkdecay}.}
    \label{fig:rad_and_bh_diagrams}
\end{figure}

In addition to experimental data, the analysis presented here makes use of \ac{mc} simulations to understand some attributes of the signal and background. MadGraph5~\cite{MG5} is used to generate signal samples at a range of masses, as well as background samples. There are two sources of background that produce $\epem$ pairs in the detector:  trident interactions in the target and \ac{wab} events.  Trident interactions are simulated with both the Bethe-Heitler and radiative diagrams  (see \cref{fig:rad_and_bh_diagrams}) including their interference terms. The \ac{wab} interactions can give a reconstructed $\epem$ pair in the detectors when the photon pair-produces either the target or the first few layers of silicon.  

The beam backgrounds, predominantly scattered single electrons, are simulated using EGS5~\cite{Hirayama:2005zm} and overlaid on all \ac{mc} samples, distributed according to the time structure of the beam to account for pileup effects. The simulation of generated samples uses \textsc{Geant4}~\cite{GEANT4:2002zbu} to model interactions with the detector, after which the detector response simulation and reconstruction are performed.

The event reconstruction follows the procedure detailed in~\cite{Adrian:2022nkt}. Briefly, energy deposits in the \ac{ecal} are grouped into clusters, with per-crystal energy corrections applied using calibration tables. These clusters are constructed by grouping high amplitude seed hits nearest and next-to-nearest neighbors. The cluster energy is then defined as the sum of energies of its constituent hits.

In the \ac{svt}, tracks are reconstructed using a combinatorial Kalman filter~\cite{Fruhwirth:1987fm} for both track finding and fitting and incorporates multiple scattering.  Each track is then propagated to the \ac{ecal} and matched to an \ac{ecal} cluster. A matched track-cluster pair is referred to as a reconstructed particle.

Pairs of oppositely charged reconstructed particles are combined to form vertex candidates. The vertex position is calculated using a global $\chi^2$ minimization algorithm~\cite{BILLOIR1992139}. Only pairs with tracks in opposite (top and bottom) halves of the detector volume are considered. 

\section{Event Selection}\label{sec:selection}
After the data samples go through reconstruction, further event selection is required to remove background \ac{sm} processes and isolate potential signal events.
This additional event selection was performed in two stages: preselection  and tight selection, as described in the sections below.

\subsection{Preselection}\label{sec:preselection}

The preselection cuts are designed to remove poorly reconstructed tracks and vertices as well as accidental $\epem$ pairs from the data sample, leaving pairs from trident and \ac{wab} events. In addition to the presence of a pair trigger, the best handle on accidental vertices are strict requirements on the differences in times between the two clusters and between track and cluster of the reconstructed particles. There are requirements on the electron energy and the $\epem$ energy sum to remove pairs where the electron is the scattered beam electron.  Additionally, well-reconstructed tracks and vertices are selected by cuts on their fit $\chi^2$s and the number of measurements on track. 
The preselection cuts are summarized in \cref{tab:PreselectionCuts}.

Each reconstructed event is then required to have exactly one of these preselected vertices.
This requirement mostly removes events in which no high-quality vertex was reconstructed;
however, this selection also eliminates pileup backgrounds and the statistical overlap of the two hit-content categories defined later for tight selections.

\begin{table}[ht] 
    \centering
    \begin{tabular}{lr}
        \toprule
        \textbf{Cut Description} & \textbf{Requirement} \\
        \midrule
        Trigger & Pair1 \\
        Track Time Relative to Trigger&  $|t_\text{trk}| \leq \SI{6}{ns} $\\
        Cluster Time Difference & $\Delta (t_{\text{clu}, e^{-}}, t_{\text{clu}, e^{+}}) \leq \SI{1.45}{ns}$ \\
        Track-Cluster Time Difference & $\Delta(t_\text{trk},t_\text{clu}) \leq \SI{4.0}{ns}$\\
        Track Quality & $\chi_\text{trk}^2/\text{n.d.f.} \leq 20.0$\\ 
        Beam Electron Cut & $p_{e^{-}} \leq \SI{1.75}{GeV}$ \\
        Minimum Hits on Track & $N_\text{hits} \geq 7$\\ 
        Unconstrained Vertex Quality & $\chi_\text{vtx}^2 \leq 20.0$  \\
        $\epem$ Momentum Sum& $p_{\text{sum}} \leq \SI{2.4}{GeV}$ \\
        \bottomrule
    \end{tabular}
    \caption{Preselection requirements for $\epem$ vertex candidates. 
    }
    \label{tab:PreselectionCuts}
\end{table}

The preselected data sample is used to optimize the displaced vertex selection cuts, described in \cref{sec:displaced-selection}. The preselected \ac{mc} sample is also used to estimate the fraction of radiative events in the data sample as a function of $\epem$ invariant mass, providing a reference for the expected signal yield that reduces the dependence on \ac{mc} modeling of experimental efficiencies as described in \cref{sec:analysis}.

\subsection{Tight Selection} \label{sec:tight-selection}
Following the preselection to produce a sample of cleanly reconstructed events with \epem vertices minimally impacted by pileup, a set of tight selections aimed specifically at sensitivity to the SIMPs signature is used to define the final event sample for the search.

\subsubsection{Signal Kinematics Selection}
In the \ac{simp} model, the \aprime decays to a stable, unobserved light dark meson $\pi_{D}$ and a heavier vector meson $V_D$. 
This shifts the signal region total momentum from near beam energy, as in the case of the nominal \aprime search, to significantly lower values; thus, a selection on the sum of the momentum magnitudes is applied:
\begin{equation}
  \psum = |\vec{p}_{e^-}|+|\vec{p}_{e^+}|.
\end{equation}
Specifically, the \ac{sr} used for the \ac{simp} search requires $\qty{1.0}{\GeV} < \psum < \qty{1.9}{\GeV}$ and the \ac{cr} used for determining the trident differential production rate and fraction of radiative tridents is
$\qty{1.9}{\GeV} < \psum < \qty{2.4}{\GeV}$.  For the \ac{simp} model considered in this work the contribution of signal events in this control region is negligible

\subsubsection{Displaced Vertex Categories}

The sources and characteristics of falsely displaced vertices depend upon the hit content of the tracks, and especially on the presence or absence of hits in the layers closest to the target. To enable the optimization of selections according to these attributes, the data is split into two mutually exclusive categories according to the hit content of the tracks.

\begin{figure}[h]
    \centering
    \resizebox{0.4\textwidth}{!}{\begin{tikzpicture}
  \drawhpsfirsttwolayers
  \node at (\targetx,+2) {L1L1};
  \node at (\targetx+1,0.) [circle,fill,inner sep=1.5pt] {};
  \draw[black,->] (\targetx+1,0.) -- (2.5,2.1) node[anchor=west] {\(e^-\)};
  \draw[black,->] (\targetx+1,0.) -- (2.5,-2) node[anchor=west] {\(e^+\)};
  \node[blue] at (\targetx,+1.5) {L1L2};
  \node[blue] at (\targetx+2.0,0.2) [circle,fill,inner sep=1.5pt] {};
  \draw[blue,->] (\targetx+2.0,0.2) -- (2.5,1.7) node[anchor=west] {\(e^-\)};
  \draw[blue,->] (\targetx+2.0,0.2) -- (2.5,-1.5) node[anchor=west] {\(e^+\)};
\end{tikzpicture}}
    \caption{Diagram showing the two mutually exclusive categories based on the track hit content within a vertex ``L1L1'' (black) and ``L1L2'' (blue).
    }
    \label{fig:L1L1_L1L2_schem}
\end{figure}
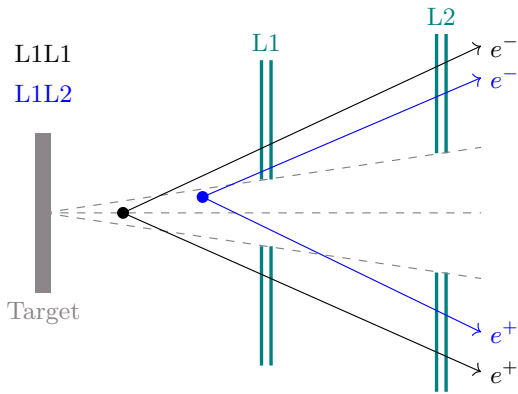

The first analysis category is called ``L1L1'', which consists of vertices where both tracks leave hits in both axial and stereo sensors in the first two tracking layers (L1 and L2).
These events have the best vertex resolution, although signal acceptance is limited to decays well upstream of L1, as depicted in \cref{fig:L1L1_L1L2_schem}. Hits in L2 are required to minimize pattern recognition errors and multiple scattering contributions in projecting tracks to the vertex.

The second analysis category is called ``L1L2'' and includes events where one track misses L1 due to a hit inefficiency or reduced acceptance due to longer lifetimes. Just as ``L1'' tracks must also have hits in L2, tracks that miss L1 are required to have hits in both L2 and L3.
The L1L2 category has $\sim50\%$ worse vertex resolution and introduces more complicated backgrounds, such as an increased rate of \ac{wab} conversions coming from the L1 material.

\subsubsection{Displaced Vertex Selection}\label{sec:displaced-selection}
The following section describes the selection procedure used to search for the displaced vertices expected in signal events. All relevant cuts are summarized in \cref{tab:tight-selection}.

Signal $\epem$ pairs should be reconstructed at a distance displaced from the target but consistent with a parent particle originating from the beamspot on the target.
This is verified by projecting a vertex candidate back towards the target at $z_\text{target}$, using the reconstructed vertex momentum.
The target-projected vertex has new coordinates \(x_\text{target}\) and \(y_\text{target}\) which can then be used to calculate a significance using the beamspot mean, $\mu_{x,y}$ and standard deviations, $\sigma_{x,y}$.
The shape, size, and position of the beamspot on the target depend on the beam conditions for a given run and are therefore characterized on a run-by-run basis. The average characteristics of the beamspot are also modeled in MC, without run dependence.
The \ac{vps}, as defined in \cref{eq:vps}, is then required to be below an optimized threshold in order to keep the vertex candidate:
\begin{equation}
\label{eq:vps}
    \text{VPS}=\sqrt{\left(\frac{x_\text{target}-\mu_{x}}{\sigma_x}\right)^2 + \left(\frac{y_\text{target}-\mu_{y}}{\sigma_y}\right)^2}~.
\end{equation}

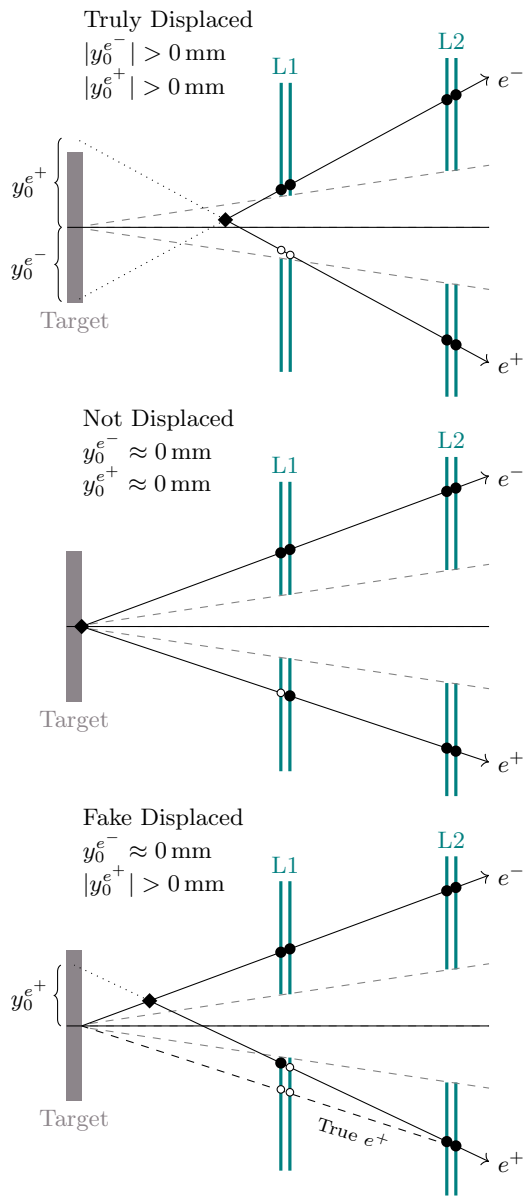
\begin{figure}[ht]
    \centering
    \resizebox{0.4\textwidth}{!}{\begin{tikzpicture}
  \drawhpsfirsttwolayers
  \draw[black] (\targetx-\targethalfthick,0) -- (2.5,0);
  \node[anchor=north west,align=left,text depth=\baselineskip] at (\targetx,+3)
    {Truly Displaced\\$|y_{0}^{e^-}| > \qty{0}{\mm}$\\$|y_{0}^{e^+}| > \qty{0}{\mm}$};
  \node at (\targetx+2,0.1) [vertex] {};
  \draw[black,->] (\targetx+2.0,0.1) -- (2.5,2) node[anchor=west] {\(e^-\)};
  \node at (\layeronex-\sensorhalfsep,\elelayeronestereoy) [hit] {};
  \node at (\layeronex+\sensorhalfsep,\elelayeroneaxialy) [hit] {};
  \node at (\layertwox-\sensorhalfsep,\elelayertwostereoy) [hit] {};
  \node at (\layertwox+\sensorhalfsep,\elelayertwoaxialy) [hit] {};
  \draw[black,->] (\targetx+2.0,0.1) -- (2.5,-1.8) node[anchor=west] {\(e^+\)};
  \node at (\layeronex-\sensorhalfsep,\poslayeronestereoy) [miss] {};
  \node at (\layeronex+\sensorhalfsep,\poslayeroneaxialy) [miss] {};
  \node at (\layertwox-\sensorhalfsep,\poslayertwostereoy) [hit] {};
  \node at (\layertwox+\sensorhalfsep,\poslayertwoaxialy) [hit] {};
  \draw[black,dotted]
    (\targetx+2,0.1) -- (\targetx,\eleyzero)
    (\targetx+2,0.1) -- (\targetx,\posyzero);
  \draw[decoration={brace,raise=5pt},decorate]
    (\targetx,\eleyzero) -- node[left=6pt] {$y_{0}^{e^-}$} (\targetx,0);
  \draw[decoration={brace,raise=5pt},decorate]
    (\targetx,0) -- node[left=6pt] {$y_{0}^{e^+}$} (\targetx,\posyzero);
\end{tikzpicture}}
    \resizebox{0.4\textwidth}{!}{\begin{tikzpicture}
  \drawhpsfirsttwolayers
  \draw[black] (\targetx-\targethalfthick,0) -- (2.5,0);
  \node[anchor=north west,align=left] at (\targetx,+3)
    {Not Displaced\\$y_0^{e^-}\approx\qty{0}{\mm}$\\$y_0^{e^+}\approx\qty{0}{\mm}$};
  \draw[black,->] (\targetx+\targethalfthick,0.0) -- (2.5,2) node[anchor=west] {\(e^-\)};
  \node at (\layeronex - \sensorhalfsep, \ndelelayeronestereoy) [hit] {};
  \node at (\layeronex + \sensorhalfsep, \ndelelayeroneaxialy) [hit] {};
  \node at (\layertwox - \sensorhalfsep, \ndelelayertwostereoy) [hit] {};
  \node at (\layertwox + \sensorhalfsep, \ndelelayertwoaxialy) [hit] {};
  \draw[black,->] (\targetx+\targethalfthick,0.0) -- (2.5,-1.8) node[anchor=west] {\(e^+\)};
  \node at (\layeronex - \sensorhalfsep, \ndposlayeronestereoy) [miss] {};
  \node at (\layeronex + \sensorhalfsep, \ndposlayeroneaxialy) [hit] {};
  \node at (\layertwox - \sensorhalfsep, \ndposlayertwostereoy) [hit] {};
  \node at (\layertwox + \sensorhalfsep, \ndposlayertwoaxialy) [hit] {};
  \node at (\targetx+\targethalfthick,0.0) [vertex] {};
  \draw[white, decoration={brace,raise=5pt},decorate]
    (\targetx,0) -- node[left=6pt] {$y_{0}^{e^+}$} (\targetx,\posyzero);
\end{tikzpicture}}
    \resizebox{0.4\textwidth}{!}{\begin{tikzpicture}
  \drawhpsfirsttwolayers
  \draw[black] (\targetx-\targethalfthick,0) -- (2.5,0);
  \node[anchor=north west,align=left] at (\targetx,+3)
    {Fake Displaced\\$y_0^{e^-}\approx\qty{0}{\mm}$\\$|y_0^{e^+}|>\qty{0}{\mm}$};
  \draw[black,->] (\targetx+\targethalfthick,0.0) -- (2.5,2) node[anchor=west] {\(e^-\)};
  \node at (\layeronex-\sensorhalfsep,\fdelelayeronestereoy) [hit] {};
  \node at (\layeronex+\sensorhalfsep,\fdelelayeroneaxialy) [hit] {};
  \node at (\layertwox-\sensorhalfsep,\fdelelayertwostereoy) [hit] {};
  \node at (\layertwox+\sensorhalfsep,\fdelelayertwoaxialy) [hit] {};
  \draw[black,->] (\fdposx, \fdposy) -- (2.5,-1.8) node[anchor=west] {\(e^+\)};
  \draw[dashed] (\layertwox,\fdposlayertwoaxialy) -- (\targetx+\targethalfthick,0.0)
    node[near start,below,sloped,font=\scriptsize] {True \(e^+\)};
  \node at (\layeronex-\sensorhalfsep,\fdbadhitstereoy) [hit] {};
  \node at (\layeronex+\sensorhalfsep,\fdbadhitaxialy) [miss] {};
  \node at (\layeronex-\sensorhalfsep,\fdtruhitstereoy) [miss] {};
  \node at (\layeronex+\sensorhalfsep,\fdtruhitaxialy) [miss] {};
  \node at (\layertwox-\sensorhalfsep,\fdposlayertwostereoy) [hit] {};
  \node at (\layertwox+\sensorhalfsep,\fdposlayertwoaxialy) [hit] {};
  \node at (\fdposx,\fdposy) [vertex] {};
  \draw[black,dotted] (\fdposx,\fdposy) -- (\targetx,\fdposyzero);
  \draw[decoration={brace,raise=5pt},decorate]
    (\targetx,0) -- node[left=6pt] {$y_{0}^{e^+}$} (\targetx,\fdposyzero);
\end{tikzpicture}}
    \caption{%
      Illustrations of the vertical track impact parameters, $y_0$, at the target
      for truly-displaced events (top), not-displaced events (middle), and fake-displaced events (bottom)
      due to scattering or reconstruction errors.}
    \label{fig:prompt_vs_displaced_diagram}
\end{figure}

Since the strip sensors of the axial (stereo) layers of the SVT are oriented with the measurement coordinate in (near) the vertical direction, the vertical impact parameter at the target, $y_0$,  has higher resolution compared
to the horizontal impact parameter and can be used to discriminate against falsely displaced vertices.
For truly displaced signal vertices, both tracks creating the vertex typically have $y_0$ far from zero. In contrast, background vertices often have one prompt track correctly reconstructed
with $y_0$ near zero, and the second track with a significant $y_0$ due to multiple scattering or mis-reconstruction.
These scenarios are depicted in \cref{fig:prompt_vs_displaced_diagram}.
This motivates selecting vertices based on requiring the minimum of the two absolute $y_0$ values to be above a mass-dependent threshold,
\begin{equation}
  \minyzero = \min(|y_{0,e^-}|,|y_{0,e^+}|)~.
\end{equation}

Finally, placing an upper limit on $\sigma_{y_0}$ for both tracks within a vertex removes some highly-displaced vertices arising from imprecisely measured tracks:
\begin{equation}
  \sigma_{y_0,\max} = \max(\sigma_{y_0,e^-},\sigma_{y_0,e^+})~.
\end{equation}

\begin{table}[h]
  \centering
    \begin{tabular}{lcc}
      \toprule
      \textbf{Selection} & \textbf{L1L1} & \textbf{L1L2} \\
      \midrule
       Momentum Sum & \multicolumn{2}{c}{$\qty{1.0}{\GeV} < \psum < \qty{1.9}{\GeV}$} \\
      From Beamspot & \ac{vps} $<2$ & \ac{vps} $<4$ \\
      Lower $y_0$ Error & -- & $\sigma_{y_0,\max} < \qty{0.4}{\mm}$ \\
      Highly Displaced & \multicolumn{2}{c}{$\minyzero > \minyzero^\mathrm{cut}(m_\text{reco})$} \\
      \bottomrule
    \end{tabular}
  \caption{%
    Summary of the final tight selection depending on hit-content category. All selection variables are explained in \cref{sec:selection}.
  }
  \label{tab:tight-selection}
\end{table}

\subsubsection{Selection Optimization}
The selections for both L1L1 and L1L2 categories are optimized independently on simulated signal samples and a \SI{10}{\%} subsample of the collected data, representing the population of background events since no sensitivity is expected at this sample size. As described previously, the minimal vertical impact parameter \minyzero of each vertex provides high discrimination power between signal and falsely displaced background events. The \minyzero is highly correlated to the reconstructed vertex position in $z$ and after optimization removes all pairs reconstructed near the target.  
 
 The final analysis is performed in \minyzero as a function of reconstructed vertex mass, $m_\text{reco}$. In this (\minyzero, $m_\text{reco}$)-space, a signal would appear as an excess of high \minyzero events in a given mass window.  Note that this differs from the approach used in~\cite{Adrian:2022nkt} where the reconstructed z-vertex position was used as the dependent variable.  

Except for \minyzero, all of the selections are optimized by keeping the signal efficiency high (at least \SI{80}{\%}) while removing background events with relatively high values of \minyzero.
While the $\sigma_{y_0,\max}$ parameter was not found to be powerful for the L1L1 category, it is helpful in removing highly-displaced background events within the L1L2 category.

Finally, the \minyzero parameter is optimized by maximizing the binomial significance of the signal yield~\cite{Cousins_2008} above the remaining background.
The signal yield calculation, described in \cref{sec:signal-yield}, is scaled up by a factor of $0.1/\epsilon$. This is done to achieve a comparable number of signal events to the background in this subsample, which is necessary in order for the optimization algorithm to work correctly.
In order to be less sensitive to statistical fluctuations and to get a smooth distribution of $\minyzero^\mathrm{cut}$ as a function of mass,
the selections chosen from this optimization were fit with a second (first) order polynomial for the L1L1 (L1L2) category.  For the L1L2 category, the first order polynomial is only used between 40-120 MeV and is taken as a contant below and above these masses. 

\begin{figure}
  \centering
  \includegraphics[width=0.95\linewidth]{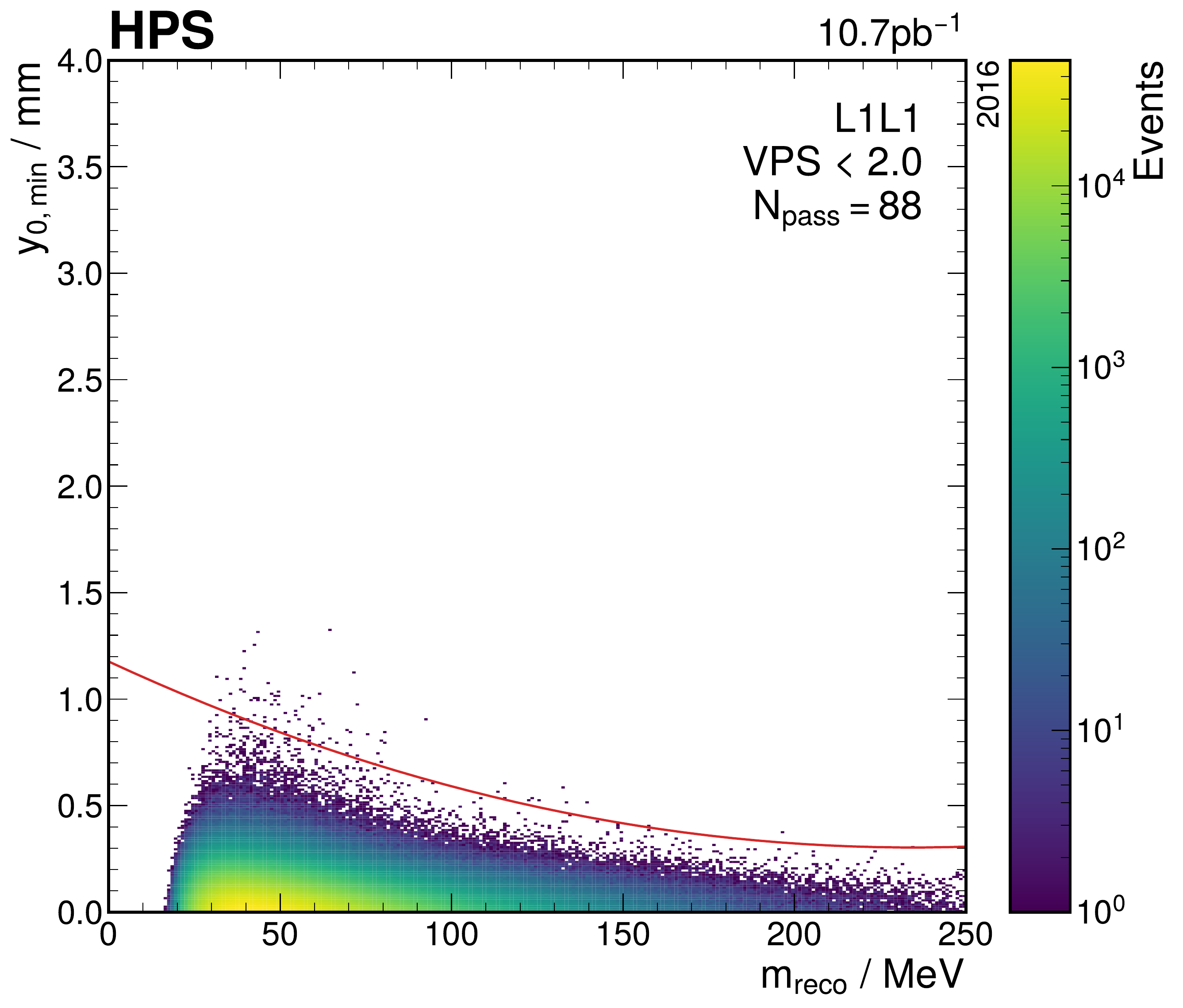}
  \includegraphics[width=0.95\linewidth]{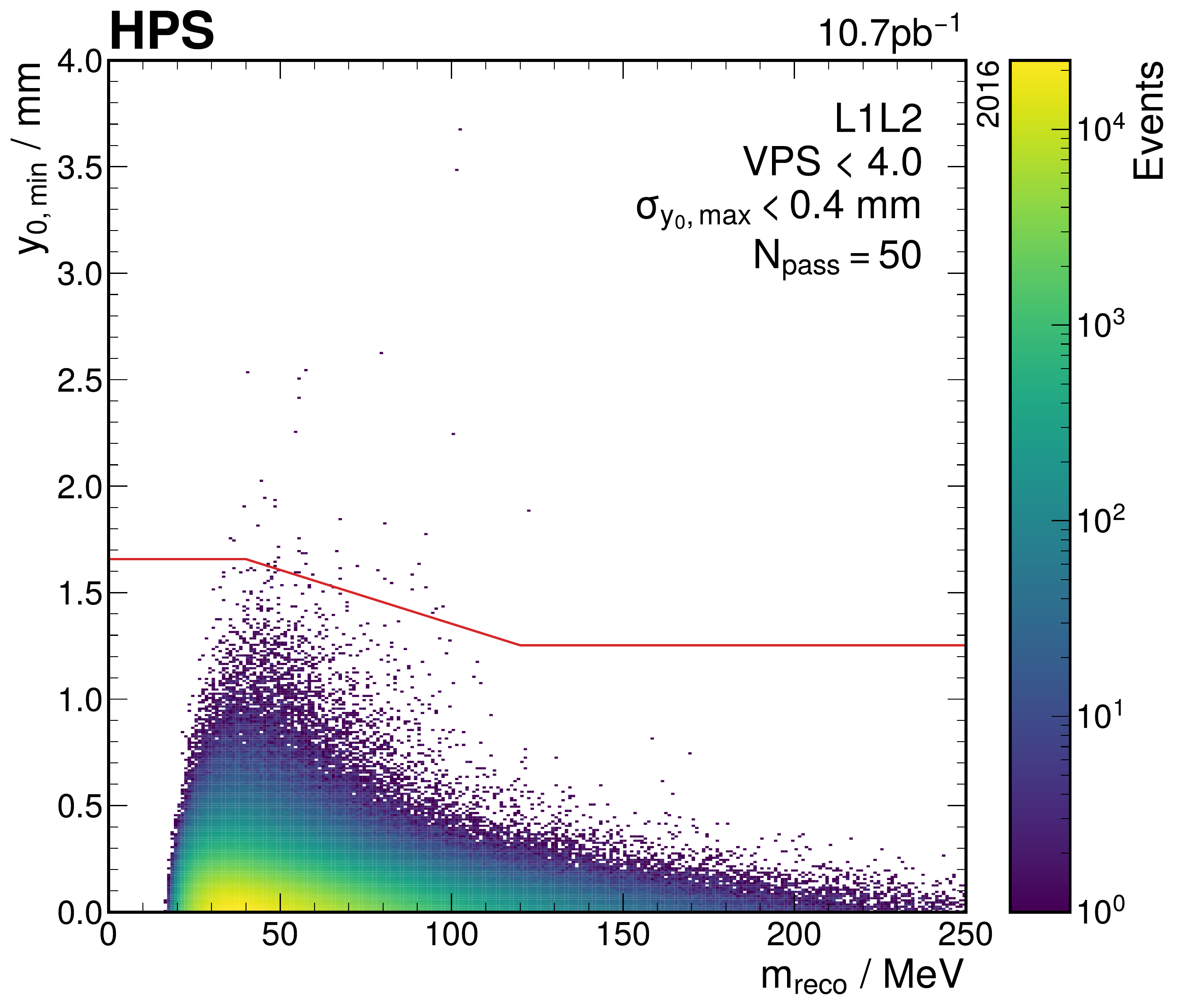}
  \caption{%
    The \minyzero distribution as a function of reconstructed invariant mass $m_\text{reco}$ with the final selection $\minyzero^\mathrm{cut}$ drawn in red for the L1L1 (L1L2) hit-content category on top (bottom). Here, a \ac{simp}-like signal would appear as an excess of high \minyzero events -- beyond $\minyzero^\mathrm{cut}$ -- within a certain mass window. The total number of events that pass the $\minyzero$ cut, $N_\text{pass}$, is noted on the plot for each category.
  }
  \label{fig:y0-cut-on-data}
\end{figure}

\cref{fig:y0-cut-on-data} shows the distributions of \minyzero as a function of $m_\text{reco}$ for the L1L1 and L1L2 hit-content categories in data after all selections have been applied. The final \minyzero cut is illustrated by the solid red line for L1L1 and L1L2 events, respectively.

\section{Data Analysis}\label{sec:analysis}
\begin{figure}[htb]
  \centering
  \includegraphics[width=0.95\linewidth]{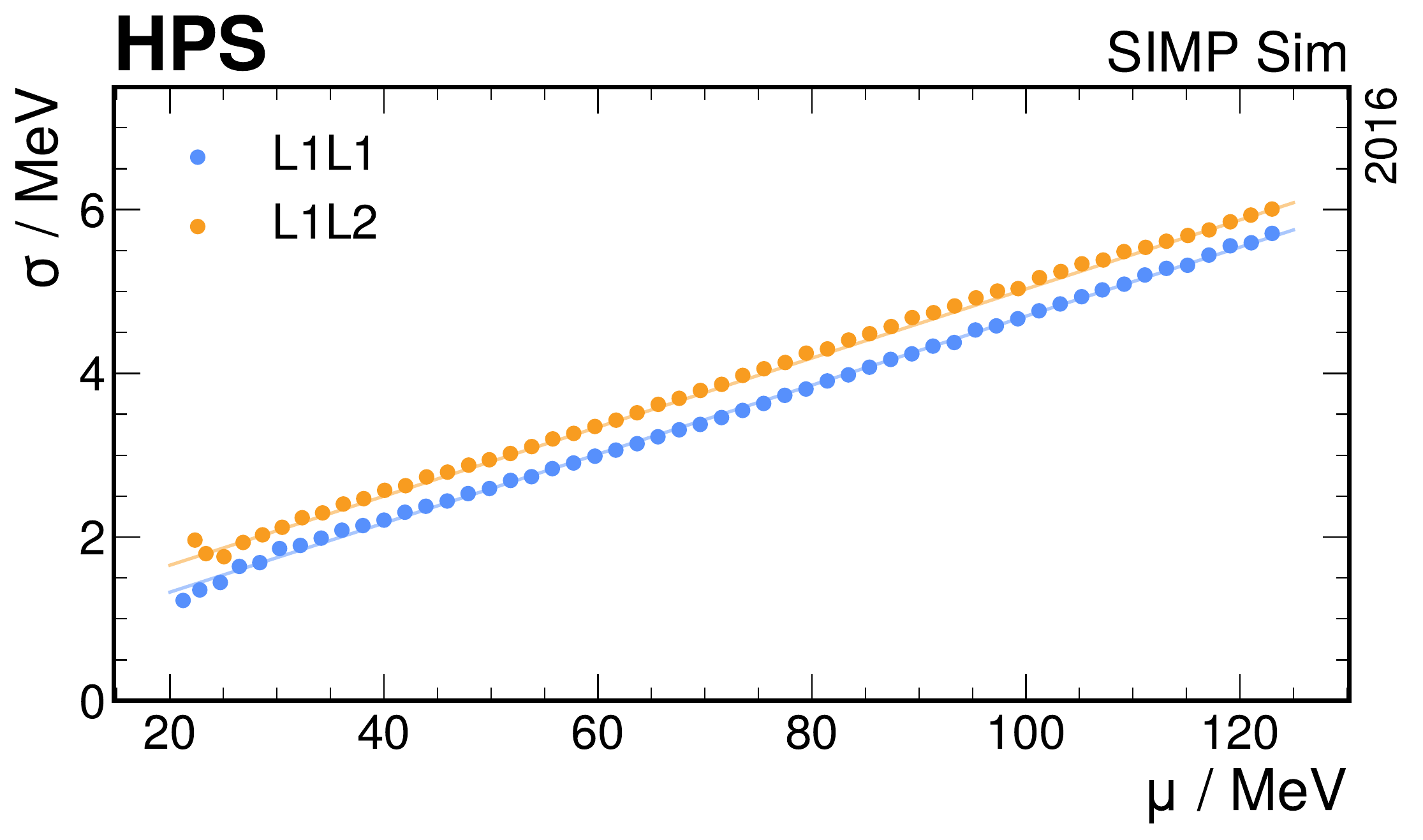}
  \caption{   
    The invariant mass resolution as estimated from Monte Carlo at various masses of $\aprime$.  The line is the result of a polynomial fit to the points and is used in the analysis.   
  }
  \label{fig:massres}
\end{figure}

This analysis searches for an excess of events in an $e^{+}e^{-}$ mass window where both tracks have large values of \minyzero, indicative of highly displaced vertices.
Additionally,  the invariant mass of the reconstructed $\epem$ pair, $m_\text{reco}$, is expected to be within a certain range of the search mass, $m_{V_{D}}$. Given that the resolution of the invariant mass peak is dominated by the detector resolution $\sigma_m$, the signal is expected to be concentrated in a region defined by:
\begin{equation}
  p_m = \frac{|m_\text{reco}-m_{V_D}|}{\sigma_m}~.
\end{equation}
Applying an upper limit on $p_m$ defines a mass window since it requires that $m_\text{reco}$ resides within a small range around $m_{V_D}$. This analysis requires $p_m < 1.5$.  The mass resolution dependence on invariant mass is shown in \cref{fig:massres}.  The mass resolution is obtained from signal MC and validated by comparing the resolution of the M\o ller scattering peak between MC and data.  More details on how the mass resolution was obtained and verified can be found in  \cite{Adrian:2022nkt}.

\subsection{Search Procedure}\label{sec:search}
\begin{figure}
  \centering
  \includegraphics[width=0.95\linewidth]{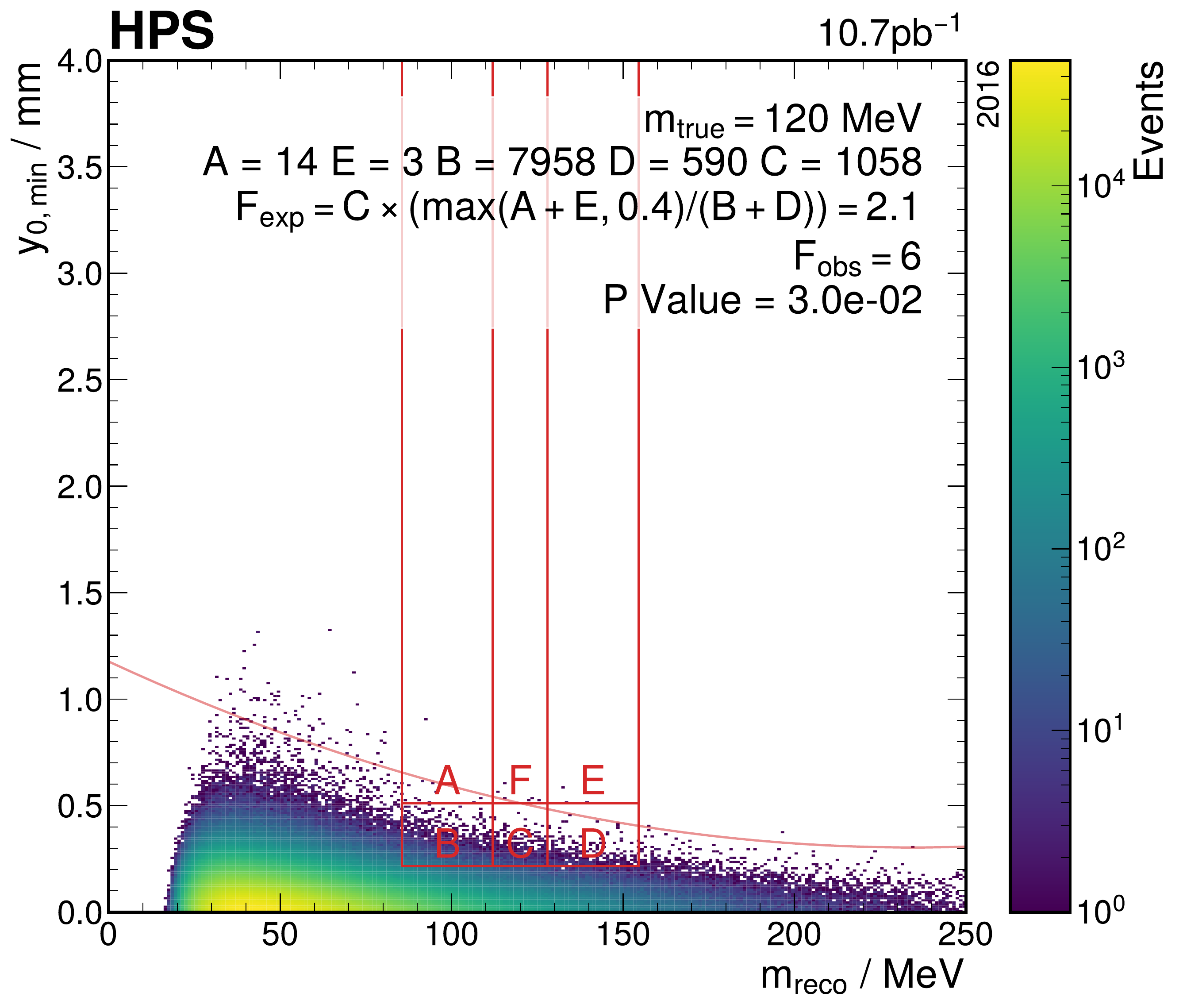}
  \caption{Example search calculation within the L1L1 channel showing the six regions
  and how the calculation is performed.}
  \label{fig:search-eg}
\end{figure}

Before applying the final selection on \minyzero, a background estimation is performed via an ABCD-like technique\cite{CDF:1990kbl,D0:1994wmk} in the (\minyzero, $m_\mathrm{reco}$)-space and compare this estimate to the observed data events to check for a signal-like excess.
The ABCD method uses sidebands to estimate the background rate in a SR. Choosing ranges in $m_\mathrm{reco}$ over which the width of the \minyzero distribution varies in a roughly linear fashion, the search space is separated into signal regions and sidebands in $m_\mathrm{reco}$ and \minyzero.
Along the $m_\mathrm{reco}$ axis, there are two sidebands -- one below and one above the signal region -- while there is one lower sideband along the \minyzero axis.
\cref{tab:abcdef-regions} gives the definition of these regions and \cref{fig:search-eg} shows an example
of these regions along with the calculation described below for the L1L1 channel.

\begin{table}[h]
  \centering
  \begin{tabular}{c|cc}
    \toprule
    \textbf{Region} & \textbf{$m_\mathrm{reco}$ Range} & \textbf{\minyzero Range} \\
    \midrule
    A & $(m_{V_D}-4.5\sigma_m,m_{V_D}-1.5\sigma_m)$ & $(y_{0,\min}^\mathrm{cut},\infty)$ \\
    B & $(m_{V_D}-4.5\sigma_m,m_{V_D}-1.5\sigma_m)$ & $(y_{0,\min}^\mathrm{floor},y_{0,\min}^\mathrm{cut})$ \\
    C & $(m_{V_D}-1.5\sigma_m,m_{V_D}+1.5\sigma_m)$ & $(y_{0,\min}^\mathrm{floor},y_{0,\min}^\mathrm{cut})$ \\
    D & $(m_{V_D}+1.5\sigma_m,m_{V_D}+4.5\sigma_m)$ & $(y_{0,\min}^\mathrm{floor},y_{0,\min}^\mathrm{cut})$ \\
    E & $(m_{V_D}+1.5\sigma_m,m_{V_D}+4.5\sigma_m)$ & $(y_{0,\min}^\mathrm{cut},\infty)$ \\
    F & $(m_{V_D}-1.5\sigma_m,m_{V_D}+1.5\sigma_m)$ & $(y_{0,\min}^\mathrm{cut},\infty)$ \\
    \bottomrule
  \end{tabular}
  \caption{
    Region definitions used in the background and signal estimation. 
    Region F is the signal region.    
  }
  \label{tab:abcdef-regions}
\end{table}

The sidebands are projected into region F to obtain the expected number of events $\mathrm{F}_\mathrm{exp}$ according to:
\begin{equation} \label{eq:f-expected}
  \mathrm{F}_\mathrm{exp} = \mathrm{C}\times\frac{\max(\mathrm{A}+\mathrm{E},0.4)}{\mathrm{B}+\mathrm{D}}~,
\end{equation}
where $\mathrm{x}$ stands for the number of events within each region.
The limiting value of $0.4$ was chosen because a Poisson mean of $0.4$ is the highest possible mean with zero observed counts being the most probable outcome.

The statistical test for excess is performed using \num{10000} toy counting experiments.
The distribution of $\mathrm{F}_\mathrm{exp}$ is constructed by sampling C and B$+$D from normal distributions
and A$+$E from a Poisson distribution, where the means of the distributions are given by the data.
This null distribution is then integrated from the observed number of events in region F
up to infinity to obtain an approximate probability that the observed number aligns with
the background prediction, which is used as the local p-value.

\begin{figure}
  \centering
  \includegraphics[width=0.95\linewidth]{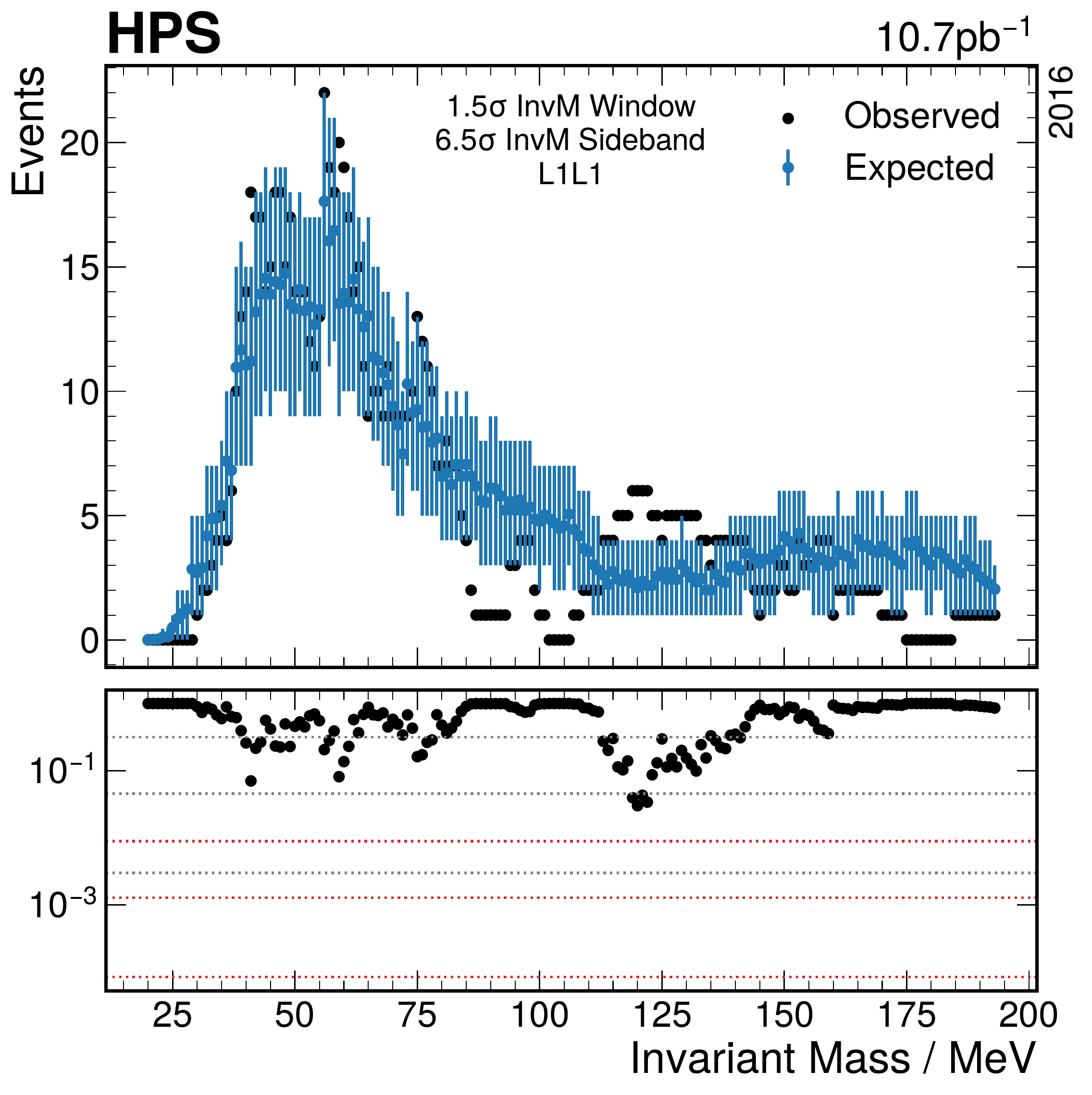}
  \includegraphics[width=0.95\linewidth]{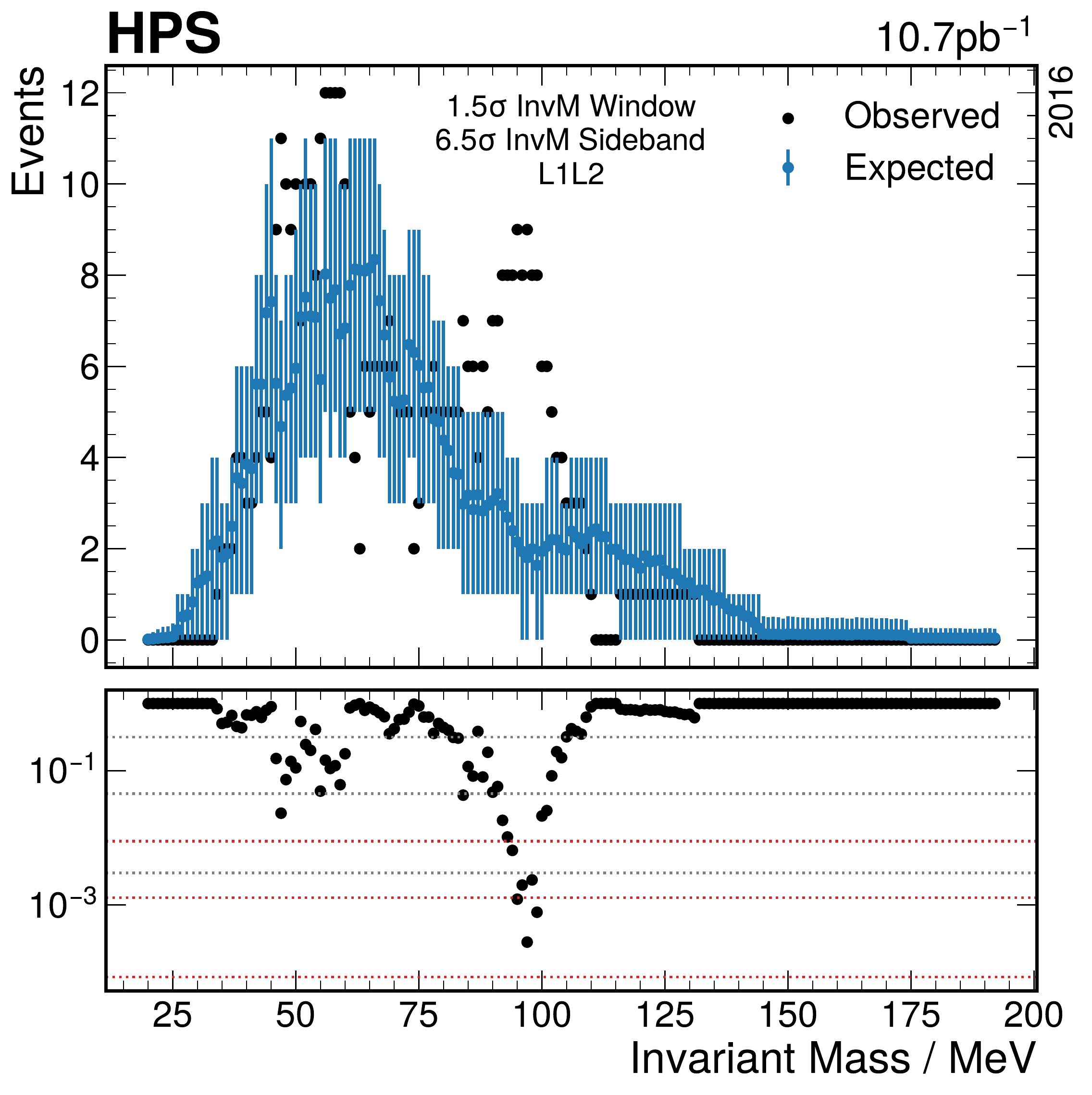}
  \caption{
    Search results for the L1L1 (L1L2) hit-content category on top (bottom).
    The gray (red) dotted lines in the lower panels are $1\sigma$, $2\sigma$,
    and $3\sigma$ local (global) significance lines.
  }
  \label{fig:search}
\end{figure}

This procedure is repeated for each mass $m_{V_D}$ in our search range, producing \cref{fig:search} showing the comparison between expected and observed event yields in region F and their corresponding p-values derived from these toy experiments.
The lowest observed p-value at $m_\mathrm{reco}=\SI{97}{MeV}$ achieves less than $3\sigma$ global significance, where the global significance is estimated by dividing the local significance by an approximate number of independent mass bins in which the search was performed. The excess only exists within the L1L2 category, supporting the conclusion that this is a normal (although rare) statistical fluctuation.

\subsection{Exclusion Procedure}\label{sec:limits} 
Without statistically significant evidence for a \ac{simp}-like signal excess an upper limit is set on the 
maximum allowed signal yield at 90\% confidence level and compared to the expectation from the model as for range of $\epsilon$ and invariant mass.  
The maximum allowed signal yield at 90\% confidence level is calculated using the \ac{oim} \cite{yellin-oim-2002}.
The limit-setting procedure on the signal yield and how that maps into exclusions in parameter space is described in the section below.  

\subsubsection{Expected Signal Yield} \label{sec:signal-yield}

The calculation for the number of $\epem$ events from $V_{D}$ decays observed in the detector is a product of the number of $\aprime$s produced in the target, the branching fraction of $\aprime\to V_{D}\pi_{D}$, and the detection efficiency of $V_{D}\to\epem$.  These calculations are detailed below. 

\begin{figure}
  \centering
  \includegraphics[width=0.95\linewidth]{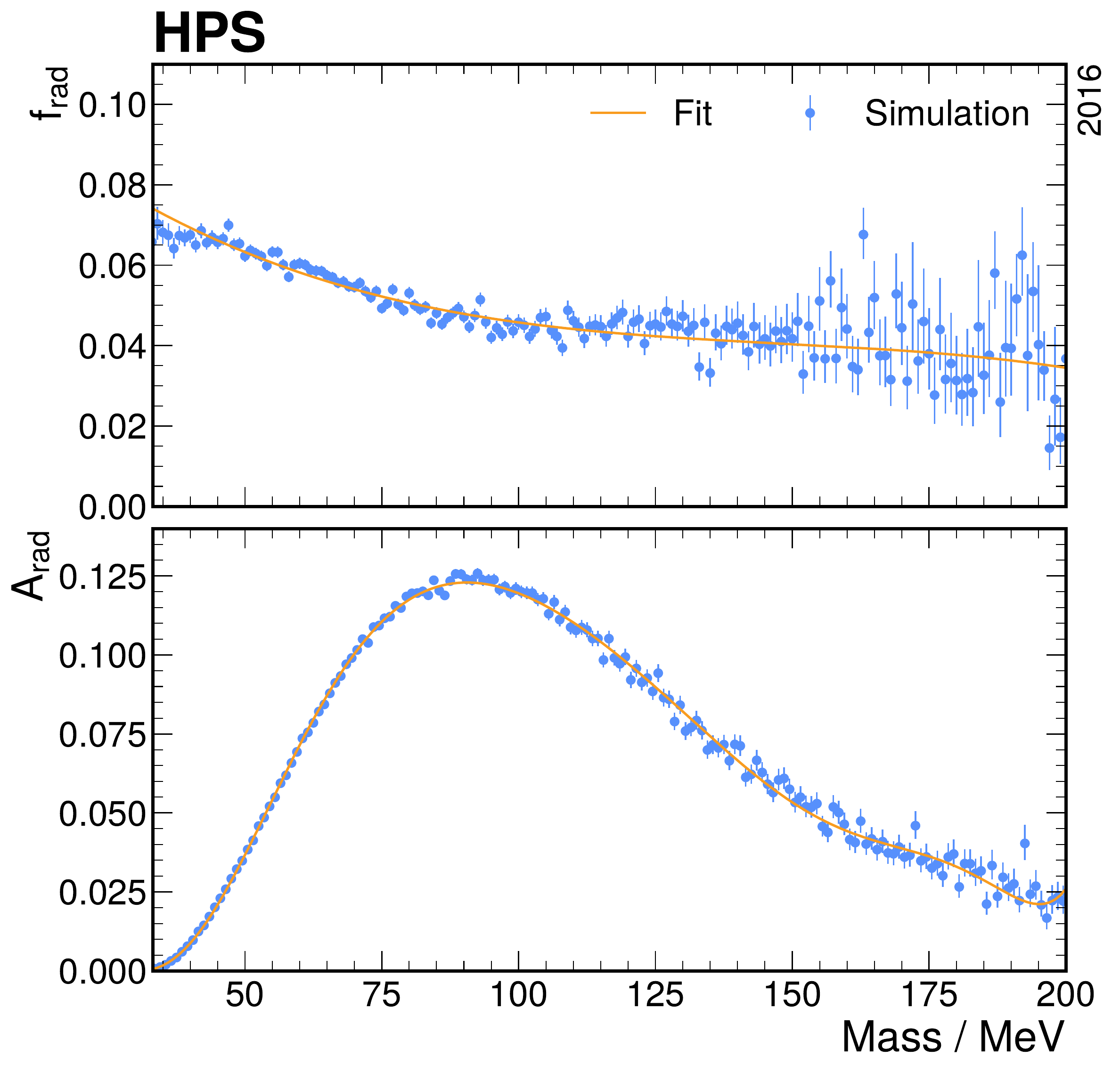}
  \caption{
      The fraction (top) and acceptance$\times$efficiency (bottom) of radiative events in our sample versus invariant mass as estimated from Monte Carlo.  The lines on the plot are from polynomial fits to the points and are what are used in the analysis.    
  }
  \label{fig:radfrac}
\end{figure}

The $\aprime$ production cross-section of a dark photon with mass $m_{\aprime}$ is related to the radiative trident production cross-section by~\cite{AprimeFixedTargetTheory}
\begin{equation} \label{eq:ap_xsec}
    \sigma_{\aprime} = \myfactors \frac{\mathrm{d}\sigma_{\gamma^*}}{\mathrm{d}m_{l^+l^-}}\Bigg|_{m_{l^+l^-} = m_{\aprime}} ~.
\end{equation}
Here, $N_\text{eff}$ is the number of available decay products (with $N_\text{eff} = 1$ since $m_\aprime < 2m_\mu$), $\alpha$ is the fine structure constant ($\alpha \approx 1/137$), and the differential cross-section is evaluated at the particular mass $m_{\aprime}$.
Multiplying both sides of \cref{eq:ap_xsec} by the integrated luminosity gives the $\aprime$ production yield given the differential radiative trident rate,
\begin{equation} \label{eq:ap_prod}
    N_{\aprime}(m_{\aprime}, \eps) = \myfactors \frac{\mathrm{d}N_{\gamma^{*}}}{\mathrm{d}m_{\aprime}}.
\end{equation}
The differential radiative trident rate in \cref{eq:ap_prod} is broken into three components as
\begin{equation} \label{eq:radtri_long}
  \resizebox{0.9 \columnwidth}{!}{$
    \frac{\mathrm{d}N_{\gamma^{*}}}{\mathrm{d}m_{\aprime}} =
      \left(\frac{\mathrm{d}N_{\gamma^{*}, \text{CR}}}{\mathrm{d}m_{\aprime}} \middle/ 
      \frac{\mathrm{d}N_\text{CR}}{\mathrm{d}m_\text{reco}}\right)
      \left(\frac{\mathrm{d}N_{\gamma^{*}}}{\mathrm{d}m_{\aprime}} \middle/
      \frac{\mathrm{d}N_{\gamma^{*}, \text{CR}}}{\mathrm{d}m_{\aprime}}\right) 
      \frac{\mathrm{d}N_\text{CR}}{\mathrm{d}m_\text{reco}}
    $}
\end{equation}
The first term in \cref{eq:radtri_long} is the radiative fraction (\radfrac), which measures the expected contribution of radiative tridents to the measured yield of \epem pairs in the \acl{cr}.  The radiative fraction has a slight dependence on invariant mass as shown in the top  of \cref{fig:radfrac}. 
The second term is the inverse of the radiative trident \accxeff, again in the \acl{cr}, referred to as the radiative acceptance (\radacc) shown in the bottom of \cref{fig:radfrac}.
The third term, $\frac{\mathrm{d}N_\text{CR}}{\mathrm{d}m_\text{reco}}$ is the measured rate of \epem pairs in the \acl{cr} and provides a means to scale the production rate to a given dataset, whether in simulation or data.

In the decay $\aprime\to V_{D}\pi_{D}$ the $V_D$ represents one of either of neutral dark vectors, $\rho_D$ and $\phi_D$, each with their production branching ratio, $BR(\aprime\to\pi_D V_D)$, and lifetime, $\Gamma(V_D\to e^+e^-)$, that are a function of $\epsilon$~\cite{simp-pheno-2018}.  The mass difference between the $\rho_D$ and $\phi_D$ is assumed to be small in this model so a search window would contain a mixture of these two dark vectors, following~\cite{simp-pheno-2018}.  
To account for this, the BR-weighted combined \accxeff for both $\rhod \rightarrow \epem$ and $\phid \rightarrow \epem$ decays is calculated as a function of the $z$-position of the $V_D$ decay. 

With $E(z)$ being the efficiency of detecting the \epem pair from a \vd decay and summing over the contributing dark vector mesons, the expected number of signal events can be estimated as:
\begin{equation}
  N_\mathrm{sig} = N_{\aprime}\int_{z_\mathrm{target}}^\infty \sum_{V_D\in\{\rho_D,\phi_D\}} D_{V_D}(z)E(z) dz
\end{equation}
where
\begin{equation}
  D_{V_D}(z) = BR(\aprime\to\pi_D V_D)
  \frac{e^{-(z-z_\mathrm{target})/(\gamma c \tau_{V_D})}}{\gamma c \tau_{V_D}}.
\end{equation}
The branching ratio $BR(\aprime\to\pi_D V_D)$ and lifetime $\tau_{V_D}$ are taken
from~\cite{simp-pheno-2018} where the lifetime explicitly depends on $m_{\aprime}$ and $\epsilon^2$.
The $V_D$ energy (and thus the relativistic $\gamma$) used in $D_{V_D}(z)$
is only distributed over a small range (within $\mathcal{O}(\qty{100}{\MeV})$)
so it is replaced with the mean $\langle\gamma\rangle$ as a simplifying assumption.

\subsubsection{Systematic Errors}
All systematic errors arising from the experiment and this analysis have been quantified individually for the two hit-content categories.
The systematic errors were found to be within $\sim\qty{1}{\%}$ of each other for both categories. The larger error of the two is used for both categories and their combination. Note that some systematic effects, which would have extended reach, were not incorporated for the purpose of obtaining a conservative estimate.
\cref{tab:systematics} summarizes the systematic uncertainties which are described in this section.

\begin{table}[h]
  \centering
  \begin{tabular}{lc}
    \toprule
    \textbf{Systematic} & \textbf{Value} \\
    \midrule
    radiative fraction & \SI{7}{\%} \\
    preselection cuts & neglected \\
    final selection cuts & neglected \\
    radiative acceptance & \\
    ~~~~~~~from pre-selection & neglected \\
    ~~~~~~~from target uncertainty & $\sim\SI{5}{\%}$ \\
    signal yield & \\
    ~~~~~~~from target uncertainty & \SI{2}{\%} \\
    ~~~~~~~from mass resolution & \SI{0.5}{\%} \\
    beamspot & neglected \\
    $\psum$ shape & $\sim 3\%$ \\
    \midrule
    total & $\sim \SI{10}{\%}$ \\
    \bottomrule
  \end{tabular}
  \caption{Summary of systematic errors considered and the values determined.
  Values marked preceded by $\sim$ are mass-dependent and
  the maximum value within the most-sensitive mass range is what is listed.}
  \label{tab:systematics}
\end{table}

The systematic error of the radiative fraction of \SI{7}{\%} is estimated from the uncertainty on the total cross sections of the different trident processes. A detailed description of this is given in~\cite{Adrian:2022nkt}.

Both preselection and final cuts have systematic errors that are found to be negligible. The difference in efficiency between data and simulated trident samples is less than a few percent for the selection variables used and is lower in the simulated background than in data.   This shift is not corrected for or included as a systematic error.
The radiative acceptance is influenced most by smearing of the pre-selection cut variables and appears to be underestimated by $\sim \qty{12}{\%}$. No correction is made for this systematic shift as this would artificially improve the sensitivity since the signal yield (and therefore the sensitivity) is inversely proportional to the radiative acceptance.

The uncertainty on the target position affects both the radiative acceptance and the signal yield. To determine the resulting systematic errors, two simulated samples with the target position offset by $\pm\qty{5}{\mm}$ were created.
This value is a conservative estimate of the uncertainty in the position of the target.
From these samples, the radiative acceptance was found to be overestimated by $\sim\qty{5}{\%}$
and the signal yield was found to be overestimated by $\qty{2}{\%}$ due to selections on
target position-dependent variables.

The width of the beamspot and the mass resolution of the detector are underestimated within the simulation
relative to the data.
In order to account for this underestimate, the resulting analysis variables were smeared accordingly. This was found to have only a small effect.
Due to a higher efficiency of events passing the \ac{vps} cut, the beamspot smearing improves the signal yield,  this effect is neglected in order to keep this exclusion estimate conservative.
The mass smearing, however, was found to decrease the signal yield by $\qty{0.5}{\%}$ which is included in the total systematic uncertainty.

Finally, the shape of the \Psum distribution is different between data and simulated background.
The effect of this systematic was determined by re-weighting events according to the ratio of the
data and simulation \Psum distributions and then re-estimating the signal yield with these new weights.
This led to a decrease in signal yield of $\sim\qty{3}{\%}$ for the most sensitive mass range, rising
to $\sim\qty{15}{\%}$ in the lower masses.

These systematic uncertainties were summed in quadrature leading to a total of $<\qty{10}{\%}$
for all but the lowest mass points evaluated (rising up to $\sim\qty{18}{\%}$).

\subsubsection{Combined Exclusion Estimates}

\begin{figure}
  \centering
  \includegraphics[width=0.95\linewidth]{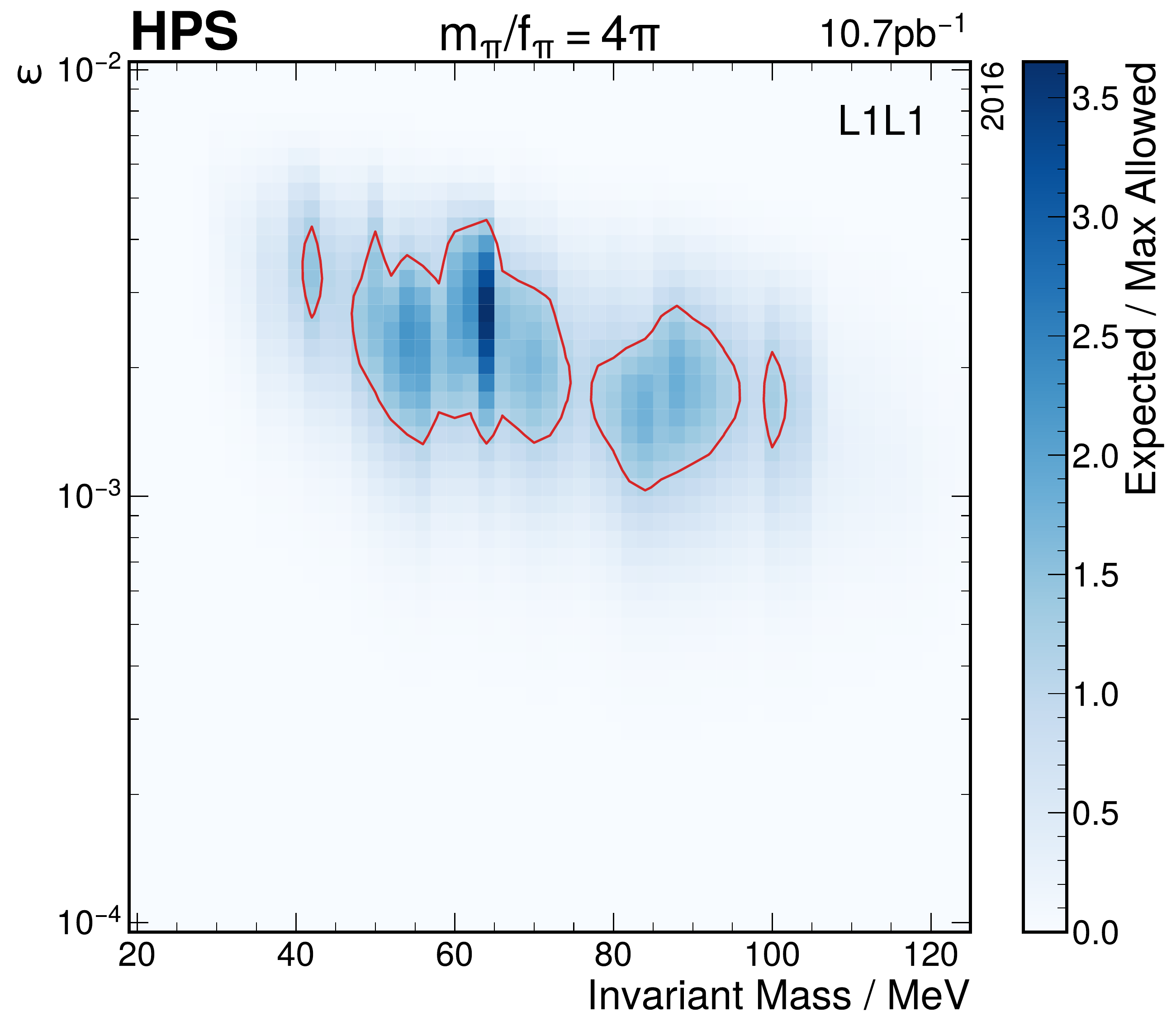}
  \includegraphics[width=0.95\linewidth]{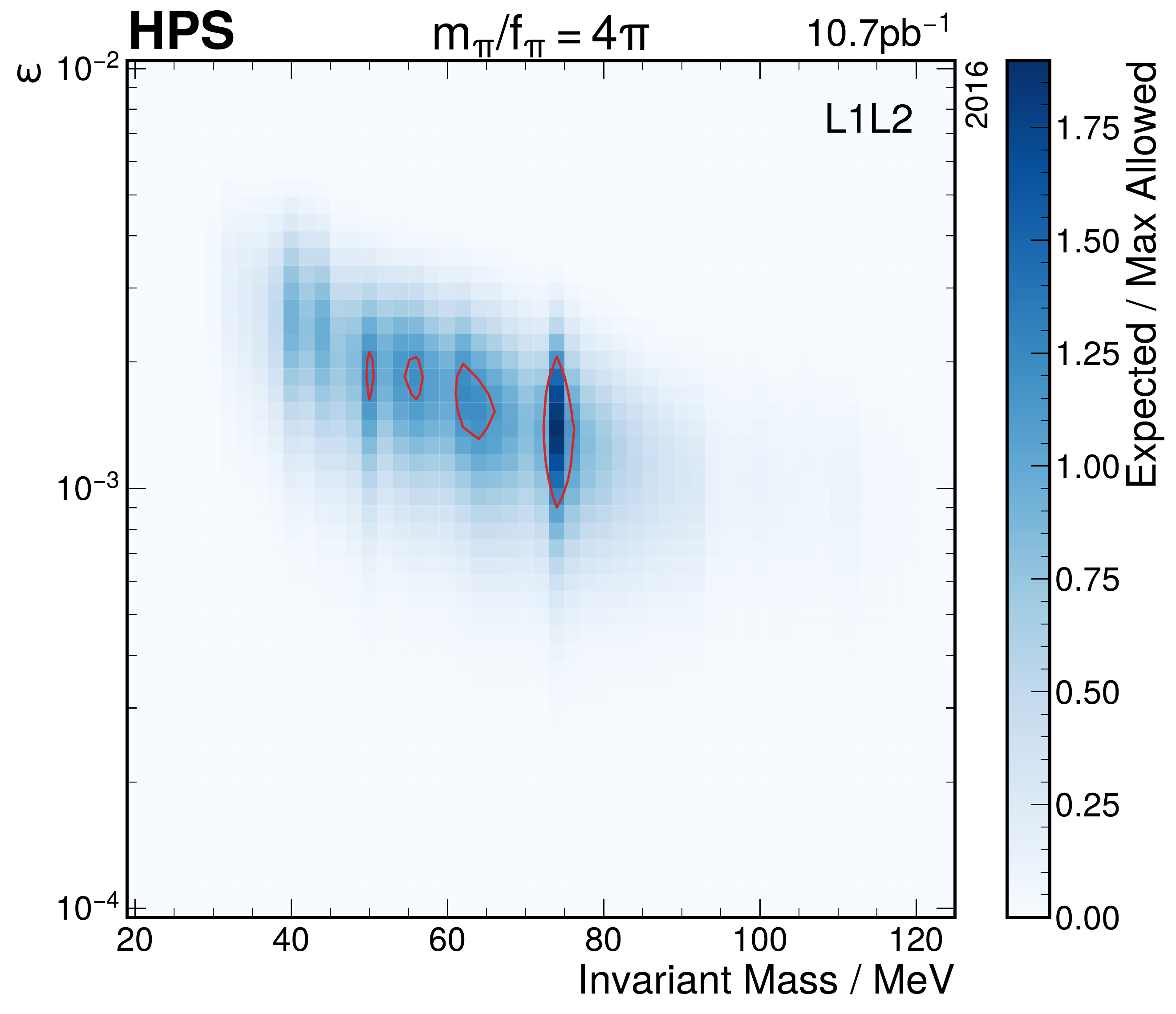}
  \caption{
    The ratio of the number of signal events expected to the maximum allowed at \qty{90}{\%} CL exclusion as a function of $\epem$ invariant mass and $\epsilon$ for L1L1 (top) and L1L2 (bottom) hit-content categories. 
    The contours outlined in red show the regions of mass-$\epsilon$ space excluded at \qty{90}{\%} CL.
  }
  \label{fig:sensitivity}
\end{figure}

\begin{figure}
  \centering
  \includegraphics[width=0.95\linewidth]{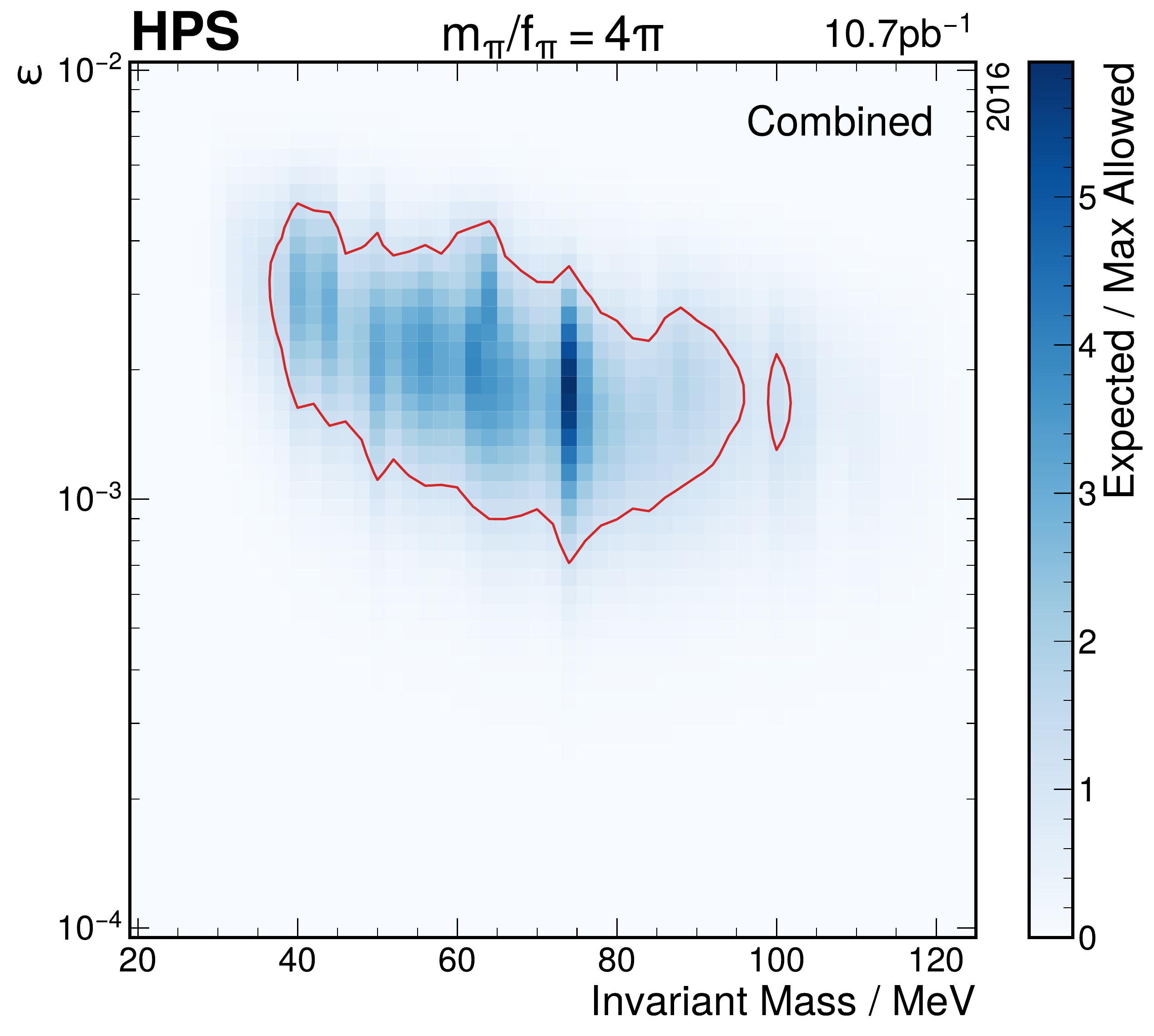}
  \caption{
   The ratio of the number of signal events expected to the maximum allowed at \qty{90}{\%} CL exclusion as a function of $\epem$ invariant mass and $\epsilon$ for the combined hit-content categories. The contours outlined in red show the regions of mass-$\epsilon$ space excluded at \qty{90}{\%} CL.     
  }
  \label{fig:combined-sensitivity}
\end{figure}

\cref{fig:sensitivity} shows the sensitivity for both hit-content categories for $m_{\pi_{D}} / f_{\pi_{D}} = 4\pi$.
The \qty{90}{\%} confidence level exclusion contours are drawn where the sensitivity equals one after being suppressed by potential systematic errors described in the previous section.
The combined sensitivity of the two categories is calculated by adding the two expected yields together and estimating the maximum allowed using the ``Minimum Limit'' combination technique
for \ac{oim} results~\cite{yellin-oim-combine-2011}.
\cref{fig:combined-sensitivity} shows the resulting sensitivity along with the combined exclusion contour, including systematic errors. Compared to the individual sensitivities of the two hit-content categories, the combined result continuously covers a broader range in invariant mass and extends to $\epsilon^2 < 10^{-6}$ which neither category reaches by itself.

The contours for $m_{\pi_{D}} / f_{\pi_{D}} = 3$, a value where the decay $\aprime\to\pi_{D}\pi_{D}$ is roughly the same as  $\aprime\to V_{D}\pi_{D}$~\cite{simp-pheno-2018}, were also calculated but found no exclusion at \qty{90}{\%} confidence level.

\section{Conclusion}\label{sec:conclusion}
\begin{figure}
  \centering
  \includegraphics[width=0.95\linewidth]{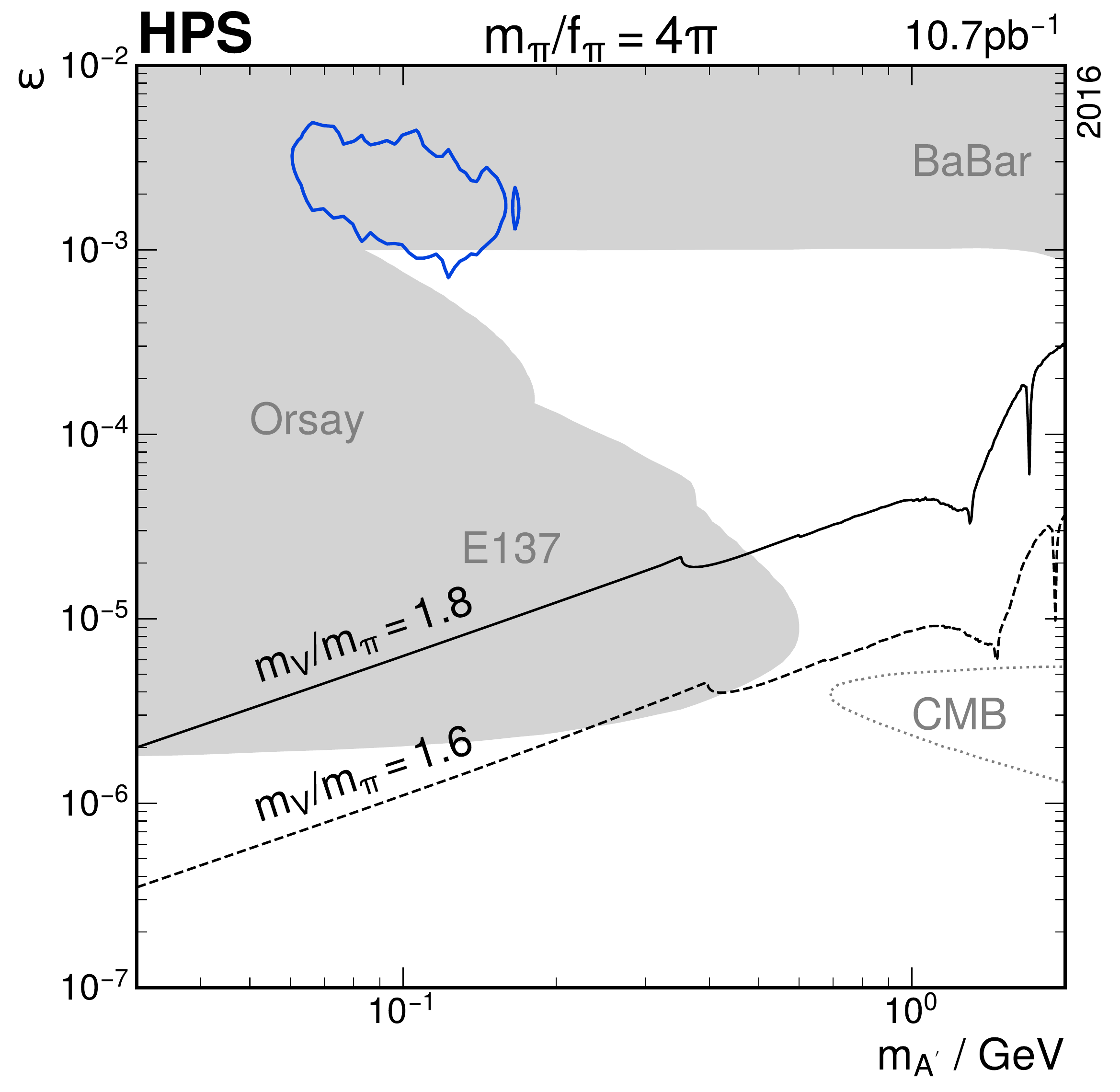}
  \caption{%
    The 90$\%$ CL exclusion contour from this analysis, with combined L1L1 and L1L2 datasets, comparisons to other experiments (gray), 
    and theoretical predictions for this model (black) \cite{simp-pheno-2018}.
  }
  \label{fig:limit-with-comparisons}
\end{figure}

In the investigated region of the \ac{simp} parameter space, couplings above $\epsilon^2 = 10^{-6}$ have been excluded by a reinterpretation~\cite{simp-pheno-2018} of BaBar~\cite{babar-2017} results. Our result, given in \cref{fig:limit-with-comparisons}, contributes to this effort by confirming the BaBar results and probing a small portion of previously unexplored \ac{simp} parameter space. Note that the lines shown in \cref{fig:limit-with-comparisons} yield the current relic abundance of DM for a given mass hierarchy; the chosen value of $m_{\pi_{D}}/f_{\pi_{D}}$ yields the highest BR of visible decays~\cite{simp-pheno-2018}, implying that the exclusion region for lower values will shrink. 

A possible extension to our analysis is given by a third hit category ``L2L2'' where both tracks miss the first tracking layer. This category also suffers from complex backgrounds and significantly worse vertex resolution, but it does have acceptance to longer lifetimes where the decay occurs downstream and neither track hit L1. The L2L2 category is particularly interesting in the context of the \ac{simp} search because there is greater acceptance for longer decay lengths. Future analyses based on the $\sim 10$  times larger 2019 and 2021 data samples could include this additional hit category.

\section{Acknowledgements}

The authors are grateful for the outstanding efforts of the Jefferson
Laboratory Accelerator Division, the Hall B engineering group, and Forest
McKinney of UC Santa Cruz in support of HPS. The research reported here is supported by the U.S.
Department of Energy Office of Science, Office of Nuclear Physics; Stanford University under Contract No. DE-AC02-76SF00515 with the U.S. Department of Energy, Office of Science, Office of High Energy Physics; the French Centre National de la 
Recherche Scientifique; 
United Kingdom's Science and Technology Facilities Council (STFC);
the Sesame project HPS@JLab funded by the French region Ile-de-France; 
and the Italian Istituto Nazionale di Fisica Nucleare. Jefferson Science
Associates, LLC, operates the Thomas Jefferson National Accelerator
Facility for the United States Department of Energy under Contract
 No. DE-AC05-060R23177.

\newpage
\bibliography{biblio}

\begin{thebibliography}{28}%
\makeatletter
\providecommand \@ifxundefined [1]{%
 \@ifx{#1\undefined}
}%
\providecommand \@ifnum [1]{%
 \ifnum #1\expandafter \@firstoftwo
 \else \expandafter \@secondoftwo
 \fi
}%
\providecommand \@ifx [1]{%
 \ifx #1\expandafter \@firstoftwo
 \else \expandafter \@secondoftwo
 \fi
}%
\providecommand \natexlab [1]{#1}%
\providecommand \enquote  [1]{``#1''}%
\providecommand \bibnamefont  [1]{#1}%
\providecommand \bibfnamefont [1]{#1}%
\providecommand \citenamefont [1]{#1}%
\providecommand \href@noop [0]{\@secondoftwo}%
\providecommand \href [0]{\begingroup \@sanitize@url \@href}%
\providecommand \@href[1]{\@@startlink{#1}\@@href}%
\providecommand \@@href[1]{\endgroup#1\@@endlink}%
\providecommand \@sanitize@url [0]{\catcode `\\12\catcode `\$12\catcode
  `\&12\catcode `\#12\catcode `\^12\catcode `\_12\catcode `\%12\relax}%
\providecommand \@@startlink[1]{}%
\providecommand \@@endlink[0]{}%
\providecommand \url  [0]{\begingroup\@sanitize@url \@url }%
\providecommand \@url [1]{\endgroup\@href {#1}{\urlprefix }}%
\providecommand \urlprefix  [0]{URL }%
\providecommand \Eprint [0]{\href }%
\providecommand \doibase [0]{https://doi.org/}%
\providecommand \selectlanguage [0]{\@gobble}%
\providecommand \bibinfo  [0]{\@secondoftwo}%
\providecommand \bibfield  [0]{\@secondoftwo}%
\providecommand \translation [1]{[#1]}%
\providecommand \BibitemOpen [0]{}%
\providecommand \bibitemStop [0]{}%
\providecommand \bibitemNoStop [0]{.\EOS\space}%
\providecommand \EOS [0]{\spacefactor3000\relax}%
\providecommand \BibitemShut  [1]{\csname bibitem#1\endcsname}%
\let\auto@bib@innerbib\@empty
\bibitem [{Hew(2012)}]{Hewett:2012ns}%
  \BibitemOpen
  \href {https://doi.org/10.2172/1042577} {\emph {\bibinfo {title}
  {{Fundamental Physics at the Intensity Frontier}}}}\ (\bibinfo {year}
  {2012})\ \Eprint {https://arxiv.org/abs/1205.2671} {arXiv:1205.2671 [hep-ex]}
  \BibitemShut {NoStop}%
\bibitem [{\citenamefont {Essig}\ \emph {et~al.}(2013)\citenamefont {Essig}
  \emph {et~al.}}]{Essig:2013lka}%
  \BibitemOpen
  \bibfield  {author} {\bibinfo {author} {\bibfnamefont {R.}~\bibnamefont
  {Essig}} \emph {et~al.},\ }\bibfield  {title} {\bibinfo {title} {{Working
  Group Report: New Light Weakly Coupled Particles}},\ }in\ \href@noop {}
  {\emph {\bibinfo {booktitle} {{Community Summer Study 2013}: {Snowmass on the
  Mississippi}}}}\ (\bibinfo {year} {2013})\ \Eprint
  {https://arxiv.org/abs/1311.0029} {arXiv:1311.0029 [hep-ph]} \BibitemShut
  {NoStop}%
\bibitem [{\citenamefont {Alexander}\ \emph {et~al.}(2016)\citenamefont
  {Alexander} \emph {et~al.}}]{Alexander:2016aln}%
  \BibitemOpen
  \bibfield  {author} {\bibinfo {author} {\bibfnamefont {J.}~\bibnamefont
  {Alexander}} \emph {et~al.},\ }\bibfield  {title} {\bibinfo {title} {{Dark
  Sectors 2016 Workshop: Community Report}}\ }(\bibinfo {year} {2016})\ \Eprint
  {https://arxiv.org/abs/1608.08632} {arXiv:1608.08632 [hep-ph]} \BibitemShut
  {NoStop}%
\bibitem [{\citenamefont {Battaglieri}\ \emph {et~al.}(2017)\citenamefont
  {Battaglieri} \emph {et~al.}}]{Battaglieri:2017aum}%
  \BibitemOpen
  \bibfield  {author} {\bibinfo {author} {\bibfnamefont {M.}~\bibnamefont
  {Battaglieri}} \emph {et~al.},\ }\bibfield  {title} {\bibinfo {title} {{US
  Cosmic Visions: New Ideas in Dark Matter 2017: Community Report}},\ }in\
  \href@noop {} {\emph {\bibinfo {booktitle} {{U.S. Cosmic Visions: New Ideas
  in Dark Matter}}}}\ (\bibinfo {year} {2017})\ \Eprint
  {https://arxiv.org/abs/1707.04591} {arXiv:1707.04591 [hep-ph]} \BibitemShut
  {NoStop}%
\bibitem [{\citenamefont {Holdom}(1986)}]{Holdom:1985ag}%
  \BibitemOpen
  \bibfield  {author} {\bibinfo {author} {\bibfnamefont {B.}~\bibnamefont
  {Holdom}},\ }\bibfield  {title} {\bibinfo {title} {{Two U(1)'s and Epsilon
  Charge Shifts}},\ }\href {https://doi.org/10.1016/0370-2693(86)91377-8}
  {\bibfield  {journal} {\bibinfo  {journal} {Phys. Lett. B}\ }\textbf
  {\bibinfo {volume} {166}},\ \bibinfo {pages} {196} (\bibinfo {year}
  {1986})}\BibitemShut {NoStop}%
\bibitem [{\citenamefont {Galison}\ and\ \citenamefont
  {Manohar}(1984)}]{Galison:1983pa}%
  \BibitemOpen
  \bibfield  {author} {\bibinfo {author} {\bibfnamefont {P.}~\bibnamefont
  {Galison}}\ and\ \bibinfo {author} {\bibfnamefont {A.}~\bibnamefont
  {Manohar}},\ }\bibfield  {title} {\bibinfo {title} {Two z's or not two
  z's?},\ }\href {https://doi.org/10.1016/0370-2693(84)91161-4} {\bibfield
  {journal} {\bibinfo  {journal} {Phys. Lett. B}\ }\textbf {\bibinfo {volume}
  {136}},\ \bibinfo {pages} {279} (\bibinfo {year} {1984})}\BibitemShut
  {NoStop}%
\bibitem [{\citenamefont {Fayet}(1990)}]{Fayet:1990wx}%
  \BibitemOpen
  \bibfield  {author} {\bibinfo {author} {\bibfnamefont {P.}~\bibnamefont
  {Fayet}},\ }\bibfield  {title} {\bibinfo {title} {{Extra U(1)'s and New
  Forces}},\ }\href {https://doi.org/10.1016/0550-3213(90)90381-M} {\bibfield
  {journal} {\bibinfo  {journal} {Nucl. Phys. B}\ }\textbf {\bibinfo {volume}
  {347}},\ \bibinfo {pages} {743} (\bibinfo {year} {1990})}\BibitemShut
  {NoStop}%
\bibitem [{\citenamefont {Bjorken}\ \emph {et~al.}(2009)\citenamefont
  {Bjorken}, \citenamefont {Essig}, \citenamefont {Schuster},\ and\
  \citenamefont {Toro}}]{AprimeFixedTargetTheory}%
  \BibitemOpen
  \bibfield  {author} {\bibinfo {author} {\bibfnamefont {J.~D.}\ \bibnamefont
  {Bjorken}}, \bibinfo {author} {\bibfnamefont {R.}~\bibnamefont {Essig}},
  \bibinfo {author} {\bibfnamefont {P.}~\bibnamefont {Schuster}},\ and\
  \bibinfo {author} {\bibfnamefont {N.}~\bibnamefont {Toro}},\ }\bibfield
  {title} {\bibinfo {title} {New fixed-target experiments to search for dark
  gauge forces},\ }\href {https://doi.org/10.1103/PhysRevD.80.075018}
  {\bibfield  {journal} {\bibinfo  {journal} {Phys. Rev. D}\ }\textbf {\bibinfo
  {volume} {80}},\ \bibinfo {pages} {075018} (\bibinfo {year}
  {2009})}\BibitemShut {NoStop}%
\bibitem [{\citenamefont {Adrian}\ \emph {et~al.}(2023)\citenamefont {Adrian}
  \emph {et~al.}}]{Adrian:2022nkt}%
  \BibitemOpen
  \bibfield  {author} {\bibinfo {author} {\bibfnamefont {P.~H.}\ \bibnamefont
  {Adrian}} \emph {et~al.},\ }\bibfield  {title} {\bibinfo {title} {{Searching
  for prompt and long-lived dark photons in electroproduced e+e- pairs with the
  heavy photon search experiment at JLab}},\ }\href
  {https://doi.org/10.1103/PhysRevD.108.012015} {\bibfield  {journal} {\bibinfo
   {journal} {Phys. Rev. D}\ }\textbf {\bibinfo {volume} {108}},\ \bibinfo
  {pages} {012015} (\bibinfo {year} {2023})},\ \Eprint
  {https://arxiv.org/abs/2212.10629} {arXiv:2212.10629 [hep-ex]} \BibitemShut
  {NoStop}%
\bibitem [{\citenamefont {Adrian}\ \emph {et~al.}(2018)\citenamefont {Adrian}
  \emph {et~al.}}]{Adrian:2018scb}%
  \BibitemOpen
  \bibfield  {author} {\bibinfo {author} {\bibfnamefont {P.~H.}\ \bibnamefont
  {Adrian}} \emph {et~al.} (\bibinfo {collaboration} {HPS}),\ }\bibfield
  {title} {\bibinfo {title} {{Search for a dark photon in electroproduced
  $e^{+}e^{-}$ pairs with the Heavy Photon Search experiment at JLab}},\ }\href
  {https://doi.org/10.1103/PhysRevD.98.091101} {\bibfield  {journal} {\bibinfo
  {journal} {Phys. Rev. D}\ }\textbf {\bibinfo {volume} {98}},\ \bibinfo
  {pages} {091101} (\bibinfo {year} {2018})},\ \Eprint
  {https://arxiv.org/abs/1807.11530} {arXiv:1807.11530 [hep-ex]} \BibitemShut
  {NoStop}%
\bibitem [{\citenamefont {Hochberg}\ \emph {et~al.}(2014)\citenamefont
  {Hochberg}, \citenamefont {Kuflik}, \citenamefont {Volansky},\ and\
  \citenamefont {Wacker}}]{PhysRevLett.113.171301}%
  \BibitemOpen
  \bibfield  {author} {\bibinfo {author} {\bibfnamefont {Y.}~\bibnamefont
  {Hochberg}}, \bibinfo {author} {\bibfnamefont {E.}~\bibnamefont {Kuflik}},
  \bibinfo {author} {\bibfnamefont {T.}~\bibnamefont {Volansky}},\ and\
  \bibinfo {author} {\bibfnamefont {J.~G.}\ \bibnamefont {Wacker}},\ }\bibfield
   {title} {\bibinfo {title} {Mechanism for thermal relic dark matter of
  strongly interacting massive particles},\ }\href
  {https://doi.org/10.1103/PhysRevLett.113.171301} {\bibfield  {journal}
  {\bibinfo  {journal} {Phys. Rev. Lett.}\ }\textbf {\bibinfo {volume} {113}},\
  \bibinfo {pages} {171301} (\bibinfo {year} {2014})}\BibitemShut {NoStop}%
\bibitem [{\citenamefont {Hochberg}\ \emph {et~al.}(2015)\citenamefont
  {Hochberg}, \citenamefont {Kuflik}, \citenamefont {Murayama}, \citenamefont
  {Volansky},\ and\ \citenamefont {Wacker}}]{PhysRevLett.115.021301}%
  \BibitemOpen
  \bibfield  {author} {\bibinfo {author} {\bibfnamefont {Y.}~\bibnamefont
  {Hochberg}}, \bibinfo {author} {\bibfnamefont {E.}~\bibnamefont {Kuflik}},
  \bibinfo {author} {\bibfnamefont {H.}~\bibnamefont {Murayama}}, \bibinfo
  {author} {\bibfnamefont {T.}~\bibnamefont {Volansky}},\ and\ \bibinfo
  {author} {\bibfnamefont {J.~G.}\ \bibnamefont {Wacker}},\ }\bibfield  {title}
  {\bibinfo {title} {Model for thermal relic dark matter of strongly
  interacting massive particles},\ }\href
  {https://doi.org/10.1103/PhysRevLett.115.021301} {\bibfield  {journal}
  {\bibinfo  {journal} {Phys. Rev. Lett.}\ }\textbf {\bibinfo {volume} {115}},\
  \bibinfo {pages} {021301} (\bibinfo {year} {2015})}\BibitemShut {NoStop}%
\bibitem [{\citenamefont {Berlin}\ \emph {et~al.}(2018)\citenamefont {Berlin},
  \citenamefont {Blinov}, \citenamefont {Gori}, \citenamefont {Schuster},\ and\
  \citenamefont {Toro}}]{simp-pheno-2018}%
  \BibitemOpen
  \bibfield  {author} {\bibinfo {author} {\bibfnamefont {A.}~\bibnamefont
  {Berlin}}, \bibinfo {author} {\bibfnamefont {N.}~\bibnamefont {Blinov}},
  \bibinfo {author} {\bibfnamefont {S.}~\bibnamefont {Gori}}, \bibinfo {author}
  {\bibfnamefont {P.}~\bibnamefont {Schuster}},\ and\ \bibinfo {author}
  {\bibfnamefont {N.}~\bibnamefont {Toro}},\ }\bibfield  {title} {\bibinfo
  {title} {Cosmology and accelerator tests of strongly interacting dark
  matter},\ }\bibfield  {journal} {\bibinfo  {journal} {Physical Review D}\
  }\textbf {\bibinfo {volume} {97}},\ \href
  {https://doi.org/10.1103/physrevd.97.055033} {10.1103/physrevd.97.055033}
  (\bibinfo {year} {2018})\BibitemShut {NoStop}%
\bibitem [{\citenamefont {Leemann}\ \emph {et~al.}(2001)\citenamefont
  {Leemann}, \citenamefont {Douglas},\ and\ \citenamefont
  {Krafft}}]{Leemann:2001dg}%
  \BibitemOpen
  \bibfield  {author} {\bibinfo {author} {\bibfnamefont {C.}~\bibnamefont
  {Leemann}}, \bibinfo {author} {\bibfnamefont {D.}~\bibnamefont {Douglas}},\
  and\ \bibinfo {author} {\bibfnamefont {G.}~\bibnamefont {Krafft}},\
  }\bibfield  {title} {\bibinfo {title} {{The Continuous Electron Beam
  Accelerator Facility: CEBAF at the Jefferson Laboratory}},\ }\href
  {https://doi.org/10.1146/annurev.nucl.51.101701.132327} {\bibfield  {journal}
  {\bibinfo  {journal} {Ann. Rev. Nucl. Part. Sci.}\ }\textbf {\bibinfo
  {volume} {51}},\ \bibinfo {pages} {413} (\bibinfo {year} {2001})}\BibitemShut
  {NoStop}%
\bibitem [{\citenamefont {Baltzell}\ \emph {et~al.}(2017)\citenamefont
  {Baltzell} \emph {et~al.}}]{HPS:2016jta}%
  \BibitemOpen
  \bibfield  {author} {\bibinfo {author} {\bibfnamefont {N.}~\bibnamefont
  {Baltzell}} \emph {et~al.} (\bibinfo {collaboration} {HPS}),\ }\bibfield
  {title} {\bibinfo {title} {{The Heavy Photon Search beamline and its
  performance}},\ }\href {https://doi.org/10.1016/j.nima.2017.03.061}
  {\bibfield  {journal} {\bibinfo  {journal} {Nucl. Instrum. Meth. A}\ }\textbf
  {\bibinfo {volume} {859}},\ \bibinfo {pages} {69} (\bibinfo {year} {2017})},\
  \Eprint {https://arxiv.org/abs/1612.07821} {arXiv:1612.07821
  [physics.ins-det]} \BibitemShut {NoStop}%
\bibitem [{\citenamefont {French}\ \emph {et~al.}(2001)\citenamefont {French}
  \emph {et~al.}}]{French:2001xb}%
  \BibitemOpen
  \bibfield  {author} {\bibinfo {author} {\bibfnamefont {M.~J.}\ \bibnamefont
  {French}} \emph {et~al.},\ }\bibfield  {title} {\bibinfo {title} {{Design and
  results from the APV25, a deep sub-micron CMOS front-end chip for the CMS
  tracker}},\ }\href {https://doi.org/10.1016/S0168-9002(01)00589-7} {\bibfield
   {journal} {\bibinfo  {journal} {Nucl. Instrum. Meth. A}\ }\textbf {\bibinfo
  {volume} {466}},\ \bibinfo {pages} {359} (\bibinfo {year}
  {2001})}\BibitemShut {NoStop}%
\bibitem [{\citenamefont {Balossino}\ \emph {et~al.}(2017)\citenamefont
  {Balossino} \emph {et~al.}}]{Balossino:2016nly}%
  \BibitemOpen
  \bibfield  {author} {\bibinfo {author} {\bibfnamefont {I.}~\bibnamefont
  {Balossino}} \emph {et~al.} (\bibinfo {collaboration} {HPS}),\ }\bibfield
  {title} {\bibinfo {title} {{The HPS electromagnetic calorimeter}},\ }\href
  {https://doi.org/10.1016/j.nima.2017.02.065} {\bibfield  {journal} {\bibinfo
  {journal} {Nucl. Instrum. Meth. A}\ }\textbf {\bibinfo {volume} {854}},\
  \bibinfo {pages} {89} (\bibinfo {year} {2017})},\ \Eprint
  {https://arxiv.org/abs/1610.04319} {arXiv:1610.04319 [physics.ins-det]}
  \BibitemShut {NoStop}%
\bibitem [{\citenamefont {Alwall}\ \emph {et~al.}(2014)\citenamefont {Alwall}
  \emph {et~al.}}]{MG5}%
  \BibitemOpen
  \bibfield  {author} {\bibinfo {author} {\bibfnamefont {J.}~\bibnamefont
  {Alwall}} \emph {et~al.},\ }\bibfield  {title} {\bibinfo {title} {The
  automated computation of tree-level and next-to-leading order differential
  cross sections, and their matching to parton shower simulations},\ }\href
  {https://doi.org/10.1007/JHEP07(2014)079} {\bibfield  {journal} {\bibinfo
  {journal} {Journal of High Energy Physics}\ }\textbf {\bibinfo {volume}
  {2014}},\ \bibinfo {pages} {79} (\bibinfo {year} {2014})}\BibitemShut
  {NoStop}%
\bibitem [{\citenamefont {Hirayama}\ \emph {et~al.}(2005)\citenamefont
  {Hirayama}, \citenamefont {Namito}, \citenamefont {Bielajew}, \citenamefont
  {Wilderman},\ and\ \citenamefont {Nelson}}]{Hirayama:2005zm}%
  \BibitemOpen
  \bibfield  {author} {\bibinfo {author} {\bibfnamefont {H.}~\bibnamefont
  {Hirayama}}, \bibinfo {author} {\bibfnamefont {Y.}~\bibnamefont {Namito}},
  \bibinfo {author} {\bibfnamefont {A.~F.}\ \bibnamefont {Bielajew}}, \bibinfo
  {author} {\bibfnamefont {S.~J.}\ \bibnamefont {Wilderman}},\ and\ \bibinfo
  {author} {\bibfnamefont {W.~R.}\ \bibnamefont {Nelson}},\ }\bibfield  {title}
  {\bibinfo {title} {{The EGS5 code system}},\ }\href@noop {} {\bibfield
  {journal} {\bibinfo  {journal} {SLAC-R-730, KEK-2005-8, KEK-REPORT-2005-8}\ }
  (\bibinfo {year} {2005})}\BibitemShut {NoStop}%
\bibitem [{\citenamefont {Agostinelli}\ \emph {et~al.}(2003)\citenamefont
  {Agostinelli} \emph {et~al.}}]{GEANT4:2002zbu}%
  \BibitemOpen
  \bibfield  {author} {\bibinfo {author} {\bibfnamefont {S.}~\bibnamefont
  {Agostinelli}} \emph {et~al.} (\bibinfo {collaboration} {GEANT4}),\
  }\bibfield  {title} {\bibinfo {title} {{GEANT4--a simulation toolkit}},\
  }\href {https://doi.org/10.1016/S0168-9002(03)01368-8} {\bibfield  {journal}
  {\bibinfo  {journal} {Nucl. Instrum. Meth. A}\ }\textbf {\bibinfo {volume}
  {506}},\ \bibinfo {pages} {250} (\bibinfo {year} {2003})}\BibitemShut
  {NoStop}%
\bibitem [{\citenamefont {Fruhwirth}(1987)}]{Fruhwirth:1987fm}%
  \BibitemOpen
  \bibfield  {author} {\bibinfo {author} {\bibfnamefont {R.}~\bibnamefont
  {Fruhwirth}},\ }\bibfield  {title} {\bibinfo {title} {{Application of Kalman
  filtering to track and vertex fitting}},\ }\href
  {https://doi.org/10.1016/0168-9002(87)90887-4} {\bibfield  {journal}
  {\bibinfo  {journal} {Nucl. Instrum. Meth. A}\ }\textbf {\bibinfo {volume}
  {262}},\ \bibinfo {pages} {444} (\bibinfo {year} {1987})}\BibitemShut
  {NoStop}%
\bibitem [{\citenamefont {Billoir}\ and\ \citenamefont
  {Qian}(1992)}]{BILLOIR1992139}%
  \BibitemOpen
  \bibfield  {author} {\bibinfo {author} {\bibfnamefont {P.}~\bibnamefont
  {Billoir}}\ and\ \bibinfo {author} {\bibfnamefont {S.}~\bibnamefont {Qian}},\
  }\bibfield  {title} {\bibinfo {title} {Fast vertex fitting with a local
  parametrization of tracks},\ }\href
  {https://doi.org/https://doi.org/10.1016/0168-9002(92)90859-3} {\bibfield
  {journal} {\bibinfo  {journal} {Nucl. Inst. Methods A}\ }\textbf {\bibinfo
  {volume} {311}},\ \bibinfo {pages} {139} (\bibinfo {year}
  {1992})}\BibitemShut {NoStop}%
\bibitem [{\citenamefont {Cousins}\ \emph {et~al.}(2008)\citenamefont
  {Cousins}, \citenamefont {Linnemann},\ and\ \citenamefont
  {Tucker}}]{Cousins_2008}%
  \BibitemOpen
  \bibfield  {author} {\bibinfo {author} {\bibfnamefont {R.~D.}\ \bibnamefont
  {Cousins}}, \bibinfo {author} {\bibfnamefont {J.~T.}\ \bibnamefont
  {Linnemann}},\ and\ \bibinfo {author} {\bibfnamefont {J.}~\bibnamefont
  {Tucker}},\ }\bibfield  {title} {\bibinfo {title} {Evaluation of three
  methods for calculating statistical significance when incorporating a
  systematic uncertainty into a test of the background-only hypothesis for a
  poisson process},\ }\href {https://doi.org/10.1016/j.nima.2008.07.086}
  {\bibfield  {journal} {\bibinfo  {journal} {Nuclear Instruments and Methods
  in Physics Research Section A: Accelerators, Spectrometers, Detectors and
  Associated Equipment}\ }\textbf {\bibinfo {volume} {595}},\ \bibinfo {pages}
  {480–501} (\bibinfo {year} {2008})}\BibitemShut {NoStop}%
\bibitem [{\citenamefont {Abe}\ \emph {et~al.}(1991)\citenamefont {Abe} \emph
  {et~al.}}]{CDF:1990kbl}%
  \BibitemOpen
  \bibfield  {author} {\bibinfo {author} {\bibfnamefont {F.}~\bibnamefont
  {Abe}} \emph {et~al.} (\bibinfo {collaboration} {CDF}),\ }\bibfield  {title}
  {\bibinfo {title} {{A Measurement of $\sigma B (W \to e \nu)$ and $\sigma B
  (Z^0 \to e^+ e-)$ in $\bar{p}p$ collisions at $\sqrt{s} = 1800$ GeV}},\
  }\href {https://doi.org/10.1103/PhysRevD.44.29} {\bibfield  {journal}
  {\bibinfo  {journal} {Phys. Rev. D}\ }\textbf {\bibinfo {volume} {44}},\
  \bibinfo {pages} {29} (\bibinfo {year} {1991})}\BibitemShut {NoStop}%
\bibitem [{\citenamefont {Abachi}\ \emph {et~al.}(1995)\citenamefont {Abachi}
  \emph {et~al.}}]{D0:1994wmk}%
  \BibitemOpen
  \bibfield  {author} {\bibinfo {author} {\bibfnamefont {S.}~\bibnamefont
  {Abachi}} \emph {et~al.} (\bibinfo {collaboration} {D0}),\ }\bibfield
  {title} {\bibinfo {title} {{Search for high mass top quark production in
  $p\bar{p}$ collisions at $\sqrt{s} = 1.8$ TeV}},\ }\href
  {https://doi.org/10.1103/PhysRevLett.74.2422} {\bibfield  {journal} {\bibinfo
   {journal} {Phys. Rev. Lett.}\ }\textbf {\bibinfo {volume} {74}},\ \bibinfo
  {pages} {2422} (\bibinfo {year} {1995})},\ \Eprint
  {https://arxiv.org/abs/hep-ex/9411001} {arXiv:hep-ex/9411001} \BibitemShut
  {NoStop}%
\bibitem [{\citenamefont {Yellin}(2002)}]{yellin-oim-2002}%
  \BibitemOpen
  \bibfield  {author} {\bibinfo {author} {\bibfnamefont {S.}~\bibnamefont
  {Yellin}},\ }\bibfield  {title} {\bibinfo {title} {Finding an upper limit in
  the presence of an unknown background},\ }\href
  {https://doi.org/10.1103/PhysRevD.66.032005} {\bibfield  {journal} {\bibinfo
  {journal} {Phys. Rev. D}\ }\textbf {\bibinfo {volume} {66}},\ \bibinfo
  {pages} {032005} (\bibinfo {year} {2002})}\BibitemShut {NoStop}%
\bibitem [{\citenamefont {Yellin}(2011)}]{yellin-oim-combine-2011}%
  \BibitemOpen
  \bibfield  {author} {\bibinfo {author} {\bibfnamefont {S.}~\bibnamefont
  {Yellin}},\ }\href {https://arxiv.org/abs/1105.2928} {\bibinfo {title} {Some
  ways of combining optimum interval upper limits}} (\bibinfo {year} {2011}),\
  \Eprint {https://arxiv.org/abs/1105.2928} {arXiv:1105.2928 [physics.data-an]}
  \BibitemShut {NoStop}%
\bibitem [{\citenamefont {Lees}\ \emph {et~al.}(2017)\citenamefont {Lees} \emph
  {et~al.}}]{babar-2017}%
  \BibitemOpen
  \bibfield  {author} {\bibinfo {author} {\bibfnamefont {J.~P.}\ \bibnamefont
  {Lees}} \emph {et~al.} (\bibinfo {collaboration} {BaBar}),\ }\bibfield
  {title} {\bibinfo {title} {{Search for Invisible Decays of a Dark Photon
  Produced in e+e- Collisions at BaBar}},\ }\bibfield  {journal} {\bibinfo
  {journal} {Physical Review Letters}\ }\textbf {\bibinfo {volume} {119}},\
  \href {https://doi.org/10.1103/physrevlett.119.131804}
  {10.1103/physrevlett.119.131804} (\bibinfo {year} {2017})\BibitemShut
  {NoStop}%
\end{thebibliography}%

\end{document}